\documentclass{llncs}

\usepackage{amsmath,amsfonts,bm,verbatim}
\usepackage{chngcntr,subcaption,colortbl,epsfig,enumitem,graphicx,hyperref,mathtools,multirow,tabularx,tikz,xcolor,xr-hyper}
\usepackage{arydshln}
\usepackage[a]{esvect}
\usepackage[misc]{ifsym}
\usepackage[all]{xy}
\usepackage{algorithm}
\usepackage[noend]{algpseudocode}
\usepackage{enumitem}
\usepackage{setspace}
\usepackage{multirow}
\usepackage{lipsum}
\usepackage{geometry}
\newcommand{\nc}{\newcommand}
\renewcommand{\v}[1]{{\bm #1}}
\nc{\nn}{\nonumber \\}
\nc{\ds}{\displaystyle}

\newtheorem{assumption}{Assumption}[section]

\begin{document}
\title{An Algorithm for Reversible Logic Circuit Synthesis\\Based on Tensor Decomposition}

\author{Hochang Lee \and Kyung Chul Jeong \and Daewan Han \and Panjin Kim\inst{1}}
\institute{
	The Affiliated Institue of ETRI,
	\quad\email{pansics@gmail.com}
}

\maketitle

\begin{abstract}
  An algorithm for reversible logic synthesis is proposed.
  The task is, for a given $n$-bit substitution map, to find a sequence of reversible logic gates that implements the map.
  The gate library adopted in this work consists of multiple-controlled Toffoli gates with $m$ control bits, where $m \in \{0, \ldots, n-1\}$. Controlled gates with large $m(>2)$ are then further decomposed into smaller gates ($m \le 2$).
  A primary goal in designing the algorithm is to reduce the number of Toffoli gates which is known to be universal.

  The main idea is to view an $n$-bit substitution map as a rank-2$n$ tensor, and to transform it such that the resulting map can be written as a tensor product of a rank-($2n-2$) tensor and the $2\times2$ identity matrix.
  It can then be seen that the transformed map acts nontrivially on $n-1$ bits only, meaning that the map to be synthesized becomes ($n-1$)-bit substitution.
  This size reduction process is iteratively applied until it reaches a tensor product of only $2\times2$ matrices.

  The time complexity of the algorithm is exponential in $n$ as most previously known heuristic algorithms for reversible logic synthesis are, but it terminates within reasonable time for not too large $n$ which may find practical uses.
  As stated earlier, our primary target is to reduce the number of Toffoli gates in the output circuit.
  Benchmark results show that the algorithm works well for hard benchmark functions, but it does not seem advantageous when the function is structured.
  As an application, the algorithm is applied to find reversible circuits for cryptographic substitution boxes, which are often required in quantum cryptanalysis.
\end{abstract}


\section{Introduction}\label{sec:1}
Beginning from the mid-twentieth century, studies on reversible computing have been motivated by several factors such as power consumption, debugging, routing, performance issues in certain cases, and so on\;\cite{RCT-review}.
The circuit-based quantum computing model is also closely related to reversible computing, where every component of the circuit works as a unitary transformation except the measurement\;\cite{bookchuang}.
As in classical logic synthesis where abstract circuit behavior is designed by using a specified set of logic gates such as $\{\rm AND, NOT\}$, quantum logic synthesis is a process of designing a logic circuit in terms of certain reversible gates.


A bijection map of the form $P_n: \{0,1\}^n \rightarrow \{0,1\}^n$ often appears as a target behavior to be synthesized in various computational problems, for example in analyzing cryptographic substitution boxes (S-box)\;\cite{aes20} or hidden weighted bit functions\;\cite{Bryant91,HWB20}, or permutation network routing\;\cite{routing}.
We can also see that such a map can be interpreted as a permutation, for example, $\sigma=(7,2,0,1,5,3,6,4)$ is a bijection map $\sigma: \{0,1\}^3 \rightarrow \{0,1\}^3$, which gives $\sigma(000) = 111$, $\sigma(001) = 010$, and so on.
Let us call it a permutation map.
Reversible logic synthesis on such maps has been studied intensively\;\cite{MMD03,SZSS10,Zak16}.
Interested readers may refer to benchmark pages\;\cite{rev-page} and\;\cite{rev-page2}, and related references therein.
It is especially a nontrivial problem to synthesize a reversible circuit for a permutation behavior that does not have an apparent structure (Section\;\ref{sec:2-2}).
Most known algorithms for unstructured permutations with $n>4$ are heuristic ones with the runtime exponential in $n$.

One of the well-known methods for synthesizing an $n$-bit permutation is to find a process of transforming $P_n$ into $\sigma_{n,{\rm id}}$, where $\sigma_{n,{\rm id}}$ is the $n$-bit identity permutation.
The reason why finding the process is equivalent to designing a reversible circuit is given in Section\;\ref{sec:2-2}.
A straightforward way to achieve the goal is to rewrite the permutation as a product of at most $2^n-1$ transpositions and to synthesize each transposition in terms of specified gates\;\cite{bookchuang}.
This naive approach is easy to implement, but the resulting circuit involves a large number of Toffoli gates.
Researchers have introduced various algorithms to improve the method, and what we have noticed is that instead of finding a direct map for identity, one may try a map that reduces the effective size of the permutation such that $P_n \mapsto P_{n-1}$ and iteratively apply it.

One of the merits of the proposed algorithm would be that it is applicable to any intermediate-size substitution map, whether or not it is structured.
A structured function (for example, some finite field arithmetic operations) can certainly be optimized through human effort, but each result can hardly be generalized to other kinds of logic behaviors.
From the perspective of productivity, it could be helpful to set a baseline by using automatic tools such as the proposed algorithm, and then look for optimizations.


This work is summarized as follows:
\begin{itemize}[noitemsep,leftmargin=*]
  \item An algorithm for reversible logic circuit synthesis is proposed.
        The algorithm is designed so that the output circuit involves as small number of Toffoli gates as possible.
        Comparisons with the previous results for benchmark functions are summarized in Table\;\ref{tab:benchmarks}.

  \item The algorithm is applied to AES\;\cite{nist-aes}, Skipjack\;\cite{skipjack}, KHAZAD\;\cite{khazad}, and DES\;\cite{des} S-boxes. 
      Except for AES, all other S-boxes do not have apparent structures where no known polynomial-time algorithm can be applied.
      Indeed, reversible circuits for these S-boxes have never been suggested prior to this work.
      It is also worth noting that the output circuits of the algorithm are always garbageless for the permutation maps.

\end{itemize}
An easy-to-use implementation of the algorithm written in Python is also available from GitHub\;\cite{code}.

The paper is organized as follows.
Section~\ref{sec:2} briefly covers the basics of reversible logic circuit synthesis and related works.
Sections~\ref{sec:3} and \ref{sec:4} are devoted to bringing out the design criteria of the algorithm.
Benchmark results and applications to cryptographic S-boxes are presented in Section~\ref{sec:5}.


\section{Background}\label{sec:2}
Logic circuit synthesis involves a functionally complete set of logic gates such as \{AND, NOT\}, which we would call a universal gate set.
In this section, we first briefly introduce widely adopted \emph{reversible} gate sets, which are called multiple-controlled Toffoli (MCT) library and NOT, CNOT, Toffoli (NCT) library\;\cite{rev-page}.


Throughout the paper, most numberings are zero-based for example `zeroth column of a matrix', except for bit positions.
In referring to the position of a bit we use one-based numberings, such as the `first' for the first bit or the `$n$-th' for the last bit.

\subsection{Gate Library}
\label{sec:2-1}
\begin{figure}[H]
	\centering
	\includegraphics[width=0.86\textwidth]{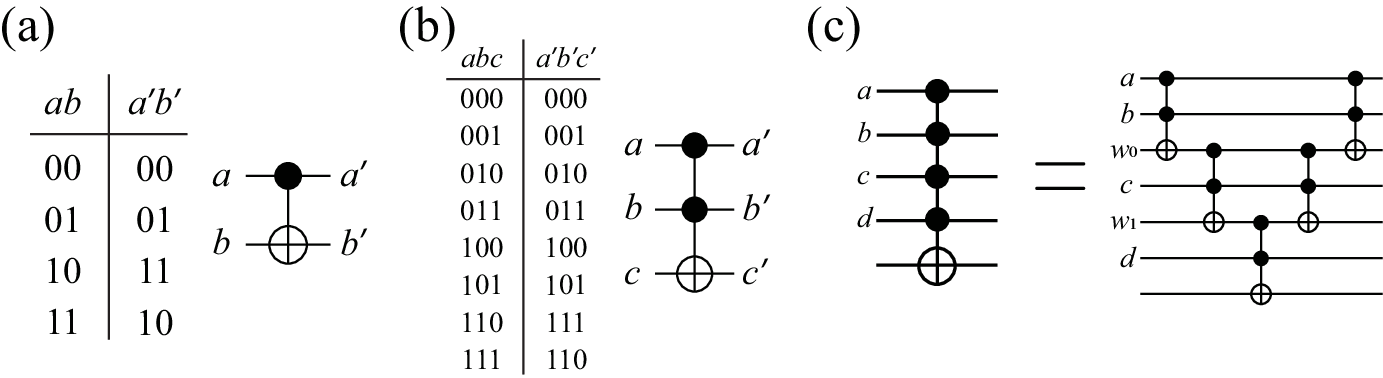}
	\caption{Truth tables and circuit symbols for (a) CNOT and (b) Toffoli gates. (c) Decomposition of $C^4\!X$ into five ($ = 2\cdot 4 -3$) Toffoli gates. Here $a,b,c,d$ are input control bits and $w_0, w_1$ are zeroed work bits.}
\label{fig:gates}
\end{figure}

Controlled-NOT, also known as CNOT, takes two input bits, one being a control and the other being a target.
Similarly, a Toffoli gate takes three input bits, two being controls and the other being a target.
Fig.\;\ref{fig:gates} illustrates CNOT and Toffoli gates which are clearly reversible.
NOT gate is already reversible, and NCT library is known to be a universal set, meaning that any reversible logic can be implemented by using these gates\;\cite{Tof80}.

Naturally, CNOT can be generalized to take $m+1$ input bits with $m$ controls and one target.
It is sometimes called an MCT gate\;\cite{rev-page}.
Let $C^m\!X$ denote such gates where $m$ is a non-negative integer.
Since NCT gates are already included in MCT library (as with $m=0,1,2$), MCT library is also a universal set.

A $C^m\!X$ gate with $m>2$ can be decomposed into a few NCT gates with various tradeoffs.
One may think of it as a conversion formula from an MCT gate to NCT gates.
For example, $C^m\!X$ can be constructed by using $2m-3$ Toffoli gates with $m-2$ zeroed work bits or by using $8m-24$ Toffoli gates with one arbitrary work bit\;\cite{barenco95}.
Fig.\;\ref{fig:gates}(c) illustrates the former formula.
Note that there can be various ways to optimize such conversion.

The number of Toffoli gates involved in a synthesized circuit will be our cost metric for a few reasons.
First, since the algorithm is for a permutation, not for a unitary operation, Toffoli gate itself is universal\;\cite{Tof80}.
We may further decompose Toffoli into even smaller gates such as T or Hadamard, but such tasks can be conveyed to independent compilation.
Secondly, we have in mind that the algorithm is applicable to quantum computing.
To our best knowledge, one of the major huddles in realizing the circuit-based quantum hardware is to implement non-Clifford gates (such as T or Toffoli) fault tolerantly.
As we are not considering synthesizing the unitaries, targeting Toffoli could be reasonable.

We say \emph{quality} of a circuit $\mathcal{C}_{\mathcal{A}}$ synthesized by an algorithm $\mathcal{A}$ is better than a circuit $\mathcal{C}_{\mathcal{B}}$ synthesized by an algorithm $\mathcal{B}$, if the number of Toffoli gates involved is smaller in $\mathcal{C}_{\mathcal{A}}$.
Note that the quality gets better as the number of Toffoli gates decreases.
An algorithm we design primarily uses MCT library, but then to measure the number of Toffoli gates, we need a conversion formula for all $C^m\!X$ gates with $m>2$.
For this purpose, we will use a simple formula described in Fig.\ref{fig:gates}(c).

\subsection{Permutation Circuit Synthesis}\label{sec:2-2}
Having an appropriate basis, an $n$-bit permutation written in one-line notation $(r_{0},r_{1}, \ldots, r_{2^{n}-1})$ can be seen as a matrix in $ \{0,1\}^{2^{n} \times 2^{n}}$.
For example for $n=3$ with the standard basis $\v b_0 = (1\; 0\; 0\; 0\; 0\; 0\; 0\; 0 )^\top$, $\v b_1 = (0\; 1\; 0\; 0\; 0\; 0\; 0\; 0 )^\top, \cdots$, a permutation $P_3 = (7, 2, 0, 1, 5, 3, 6, 4)$ can be written by a truth table and by a matrix.

\begin{align}\label{eq:mat-example}
\small
\setstretch{0.9}
\begin{tabular}{c|c}
\text{in} & \text{out} \tabularnewline\hline
000     &   111      \tabularnewline
001  	  &   010      \tabularnewline
010  	  &   000      \tabularnewline
011     &   001      \tabularnewline
100  	  &   101      \tabularnewline
101  	  &   011      \tabularnewline
110     &   110      \tabularnewline
111  	  &   100      \tabularnewline
\end{tabular},
\quad
P_3 =
\left(
{
	\arraycolsep=3pt
	\begin{array}{cccccccc}
	0 & 0 & 1 & 0 & 0 & 0 & 0 & 0 \\
	0 & 0 & 0 & 1 & 0 & 0 & 0 & 0 \\
	0 & 1 & 0 & 0 & 0 & 0 & 0 & 0 \\
	0 & 0 & 0 & 0 & 0 & 1 & 0 & 0 \\
	0 & 0 & 0 & 0 & 0 & 0 & 0 & 1 \\
	0 & 0 & 0 & 0 & 1 & 0 & 0 & 0 \\
	0 & 0 & 0 & 0 & 0 & 0 & 1 & 0 \\
	1 & 0 & 0 & 0 & 0 & 0 & 0 & 0
	\end{array}
}
\right),
\end{align}
where we have used the same symbol $P_n$ interchangeably to denote the one-line notation of an $n$-bit permutation and the corresponding matrix.
Having this in mind, we define permutation circuit synthesis as follows:

\begin{definition}
	For an $n$-bit permutation $P_n$, a finite universal gate set $G$, and a cost metric on a set, permutation circuit synthesis is to find a finite ordered set $R = \{g_i \;|\; g_i \in G \}$ such that $P_n \cdot \left(g_1 \cdot g_2 \cdot \cdots \cdot g_{|R|} \right) = I_{2^n} $, minimizing the cost on $R$. 
\end{definition}
Let $X_i$ denote a NOT gate acting on $i$-th bit, $C\!X_{ij}$ denote a CNOT gate controlled by $i$-th bit acting on $j$-th, $C\! X_{ijk}$ denote a Toffoli gate controlled by $i$- and $j$-th bits acting on $k$-th bit.
One can verify that $P_3 \cdot C\!X_{13} \cdot C\!X_{312} \cdot X_{2} \cdot C\!X_{23} \cdot C\!X_{231} = I_{2^3}$ (gate action is described for example in Eq.\;\eqref{eq:gate-actions}), and thus $C\!X_{231} \cdot C\!X_{23} \cdot X_{2} \cdot C\!X_{312} \cdot C\!X_{13} = P_3$ since NCT gates are involutory ($A=A^{-1}, A\in\text{NCT}$).

The permutation circuit synthesis is closely related with symmetric groups.
A set of $n$-bit permutations form the symmetric group ${\mathcal{S}}_{2^n}$, where $|{\mathcal{S}}_{2^n}| = 2^n !$.

\subsection{Related Works}\label{sec:2-3}
Study of algorithms for reversible logic circuit synthesis has been an active area of research over the last decades.
One question that can be asked when one sees, for example the synthesis of $P_3$ as above, is whether the obtained circuit is optimal.
In fact, the optimality of small-bit permutations (with a certain metric), or even larger permutations with linear structures, have been studied\;\cite{lowerbound,perm06,perm07,perm08,perm10,perm12,optimal-clif-linear}.
The algorithms \emph{search} the target circuit such that its minimal cost can be guaranteed, for example by exhaustive search or meet-in-the-middle method.

From a theoretical point of view, a natural question is how small the synthesized circuit can be for given bijective function $f:\{0,1\}^n \rightarrow \{0,1\}^n$.
Let $G$ be a gate library and let $K_n$ be a set of $n$-bit permutations that can be synthesized by the gates in $G$.
It has been proven by\;\cite{lowerbound} that the worst case $n$-bit permutation can be synthesized by $\Omega(|K_n| / \log |G|)$ gate-length circuit on $n$ wires.
If we choose $G=\rm NCT$, then the lower bound reads $\Omega(n 2^n / \log n)$.
The paper has also shown that for a linear function, the worst-case synthesis can be as short as $\Omega(n^2 / \log n)$ gate-length circuit.
The asymptotically optimal algorithm for linear functions did come out later\;\cite{optimal-linear} with $O(n^2 / \log n)$ gate-length output.

Apart from the worst case, when a permutation is structured, it is often possible to find an efficient circuit by exploiting its structure.
Here by `structured' we mean $r_i$ from the previous subsection is efficiently computable such that $r_i = g(i)$ with some function $g$.
If such a function is known, one may design an algorithm that computes $g$ reversibly.
Examples include arithmetic operations such as squaring in $\mathbb{F}_2^n$, or cryptographic S-boxes\;\cite{nist-aes,aes16}.


On the other hand, when an $n$-bit permutation shows no apparent structure with $n$ being larger than 4, then we are left with options using heuristic algorithms\;\cite{RCT-review} or reversible lookup table methods\;\cite{lut13}.
A reversible lookup table implements each $r_i$ for corresponding $i$ one by one, reversibly.
Although straightforward and possibly not less efficient than heuristic algorithms asymptotically, the reversible lookup table method is not preferred over heuristic algorithms for the problems at hand to our consideration.


There seems no clear criterion in classifying the known heuristic algorithms, but we briefly introduce a few well-known branches.
When the synthesis is viewed as a search problem, since the search space grows exponentially as the number of bits increases, one may try \emph{heuristic search}.
For example the greedy method can be used as in\;\cite{gupta06}, where priority-based tree search is applied with pruning.
\emph{Transformation-based} algorithms find an ordered set of gates that the sequential multiplications on the given permutation results in the identity function\;\cite{MMD03}, as we have seen in Section\;\ref{sec:2-2}.
One obvious way is to write down the permutation as a composition of transpositions and implement each transposition by reversible gates.
\emph{Cycle-based} algorithms try to decompose the given permutation in terms of smaller cycles and then synthesize each cycle\;\cite{lowerbound,SZSS10}.
Not being entirely independent of the above methods, but a number of approaches rely on \emph{functional representation} of the given permutation, looking for reversible circuits corresponding to each part of the representation.
Examples include decision diagram\;\cite{Kerntopf04,QMDD}, exclusive-or-sum-of-product\;\cite{Sasao93,fazel07}, and positive-polarity Reed-Muller expansion\;\cite{AgrawalJ04}.

The algorithm we propose recursively looks for the decomposition of smaller tensors (Section\;\ref{sec:3}).
A similar notion has been investigated by\;\cite{VosR08}, utilizing Young subgroups.
Their algorithm makes use of the fact that for any $a \in {\mathcal{S}}_{2^n}$, there exists a decomposition $a = h_1 \cdot v \cdot h_2$ where $h_i$ and $v$ are elements of the certain Young subgroups.
By finding efficiently implementable $h_1$ and $h_2$, and then by recursively applying the method to $v$, one obtains the reversible circuit.
Therefore, the recursive procedure is always applied to the center element of the decomposition, resulting in a kind of two-way construction (gates are applied from the front ($h_1$) and from the end ($h_2)$).
The proposed algorithm in this work is different from\;\cite{VosR08}, as it is a kind of one-way construction (Section\;\ref{sec:3-1}).
Perhaps there could be a connection between\;\cite{VosR08} and ours, for example there is a chance that the tensor decomposition we look for in each recursion is related to Young subgroups, but it is inconclusive in our analysis.
One advantage of the proposed algorithm is that since it recursively reduces the problem size, one bit at a time becomes free.
More discussion is given in Section\;\ref{sec:3-1}.

Due to the heuristic nature of the algorithms, it is not easy to analyze the pros and cons of each algorithm.
Therefore, benchmark functions are often synthesized by the newly proposed algorithms for the comparisons\;\cite{rev-page,rev-page2}.
The results are given in Section\;\ref{sec:5} and\;\cite{code}.


A slightly off-topic but related problem is synthesis of unitary gates (Chapter 4, \;\cite{bookchuang}).
Similar to reversible logic synthesis, various approaches are examined for general unitaries\;\cite{AMMR13,barenco95,whaley04}, or structured (Clifford) ones\;\cite{stabilizer,optimal-clif-linear,clif-6}.
For Clifford circuits, 6-bit functions are synthesized in near optimal ways\;\cite{clif-6}.


\section{Basic Idea}\label{sec:3}
The basic strategy this work has taken is to reduce the effective size of a matrix at each step by looking for a size reduction process, leaving one bit completely excluded from subsequent steps.
The size reduction idea naturally results in an algorithm for permutation circuit synthesis under an assumption.
The assumption will be lifted in Section\;\ref{sec:4} , leading to an algorithm that works for any permutation.

\subsection{Matrix Size Reduction}\label{sec:3-1}
Reversible gates acting on $n$ bits can be viewed as rank-$2n$ tensors\;\cite{bookJohn}.
Among them, there exist certain gates of which decomposition as a tensor product of smaller tensors can be found.
For example, Walsh-Hadamard transformation acting on two bits can be represented as a $4\times4$ matrix (or equivalently rank-$4$ tensor), which can also be decomposed into two $2\times2$ matrices such as

\begin{align}\label{eq:W-H}
H^{\otimes 2}
= \frac{1}{2}
\left(
\begin{array}{cccc}
1 &  1 &  1 &  1 \\
1 & -1 &  1 & -1 \\
1 &  1 & -1 & -1 \\
1 & -1 & -1 &  1 \\
\end{array}
\right)
= \frac{1}{\sqrt{2}}
\left(
\begin{array}{cc}
1 &  1 \\
1 & -1 \\
\end{array}
\right)  \otimes
\frac{1}{\sqrt{2}}
\left(
\begin{array}{cc}
1 &  1 \\
1 & -1 \\
\end{array}
\right) \enspace .
\end{align}
However, CNOT cannot be written as a simple tensor product of two smaller tensors.

Now consider a permutation $P_n$ that can be written as $P_{n-1} \otimes I_{2}$.
It means the gate $P_n$ acts nontrivially on $n-1$ bits only, effectively leaving one bit irrelevant from the operation.
It is clear that if one is able to find a procedure for transforming an arbitrary rank-$2n$ tensor into one that can be decomposed into one rank-($2n-2$) tensor and one $2\times2$ identity matrix, the circuit can be synthesized by iterating the procedure at most $n-1$ times.

A somewhat related notion has been examined in synthesizing unitary matrices\;\cite{BVMS04,SAM11}, but apart from the use of the notion \emph{block} (Section\;\ref{sec:3-2}), the design criteria are quite different.
In a nutshell, their method is to divide a large circuit into several smaller circuits and each one is recursively divided into even smaller ones, involving gate operations to every bit all the way through to the end.
On the contrary, each time the size of the permutation gets smaller in our method, one bit becomes completely irrelevant from the circuit.



\subsection{Definitions and Conventions}\label{sec:3-2}
In $n$-bit permutation circuit synthesis, a gate set we use consists of $C^{m}\!X$ gates, where the number of control bits ranges from 0 to $n-1$.

Dealing with a permutation is probably best comprehensible in matrix representation as in Eq.\,(\ref{eq:mat-example}), with an obvious downside that it is inconvenient to write down.
One-line notation is thus frequently adopted throughout the paper.
For example, the action of $X_1$, $C\!X_{21}$, and $C\!X_{312}$ on $(7,2,0,1,5,3,6,4)$ reads
\begin{align}\label{eq:gate-actions}
&\overset{X_1}{~~\longmapsto~~} \left(\colorbox[rgb]{0.7,0.7,0.7}{\makebox(30,6){$5,3,6,4$}},\colorbox[rgb]{0.7,0.7,0.7}{\makebox(30,6){$7,2,0,1$}}\right) ,\nonumber \\[1pt]
(7,2,0,1,5,3,6,4) &\overset{C\!X_{21}}{~~\longmapsto~~} \left(7,2,\colorbox[rgb]{0.7,0.7,0.7}{\makebox(11,6){$6,4$}},5,3,\colorbox[rgb]{0.7,0.7,0.7}{\makebox(11,6){$0,1$}}\right),  \\[1pt]
&\overset{C\!X_{312}}{~~\longmapsto~~} \left(7,2,0,1,5,\colorbox[rgb]{0.7,0.7,0.7}{\makebox(1.5,6){$4$}},6,\colorbox[rgb]{0.7,0.7,0.7}{\makebox(1.5,6){$3$}}\right).\nonumber
\end{align}
When a permutation is written by $P_n=(r_0, r_1, ..., r_{2^n -1})$, the subscripts $i$ and integers $r_i$ are understood as \emph{column numbers} and \emph{row numbers}, respectively, in which nonzero values reside in the matrix representation.
For example, $r_0=7$ means 1 is located at the 0th column and the 7th row in the matrix as in Eq.\;(\ref{eq:mat-example}).
Using these notions, a permutation $P_n$ can be viewed as a function of a column number,
\begin{align*}
P_n:\{0,1,\cdots,2^n -1\} \rightarrow&\, \{0,1,\cdots,2^n -1\} , \nn
i ~~~~~\mapsto&~~~~~ r_i .
\end{align*}
Column numbers will frequently be read as $n$-bit binary strings.
Binary strings will be denoted by vector notation when necessary.
In writing an integer $x$ as an $n$-bit binary string $\v x\in \{0,1\}^n$, we let $x_i\in\{0,1\}$ denote the $i$-th bit of $\v x$ for $1 \leq i \leq n$.

One way to understand the action of a logic gate is to think of it as an operator that exchanges column numbers.
When a gate is applied, it first reads column numbers as $n$-bit binary strings.
Among the strings (columns), some must meet the condition for the activation of the gate.
The row numbers that reside in these columns are swapped appropriately.
For example in Eq.\;(\ref{eq:gate-actions}), each column number is read as $000,001,\cdots,111$ which are occupied by row numbers $7,2,\cdots,4$, respectively.
The gate $C\!X_{132}$ is activated when the value of the first and the third bits are both 1, which correspond to the 5th (101) and the 7th (111) columns.
Row numbers 3 and 4 which preoccupied these positions are then swapped by the gate.

In addition to $X_i$, $C\!X_{ij}$, and $C\!X_{ijk}$ gates defined in Section\;\ref{sec:2-2}, we introduce a way to specify control and target bits of general $C^m\!X$ gates with nonnegative integer $m$.
Let $\mathcal{I}$ be a subset of $\{1,\cdots,n\}$.
In denoting a controlled gate with an arbitrary number of control bits, an index set $\mathcal{I}$ will be used such that $C\!X_{\mathcal{I}:k}$ has $|\mathcal{I}|$ control bits specified by elements in $\mathcal{I}$ and targets $k$-th bit.

The following definition for \emph{block} is the central notion in this work, and it is recommended to refer to the example in Eq.\,(\ref{eq:block-example}) before comprehending the formal definition.
%
\begin{definition}
	\label{def:block}
	For an $n$-bit permutation $\left(r_{0}, \ldots, r_{2^{n}-1}\right)$, a pair of numbers $r_{2i}$ and $r_{2i+1}$ is defined as an even (odd) block if $r_{2i+1}-r_{2i} = 1\;(-1)$, where $i\in \{0,1,...,2^{n-1}-1\}$ is called a block-wise position.
\end{definition}
It is unlikely that the number of blocks found in an unstructured permutation is large.\footnote{It is related with Hat matching (or Hat check) problem. For an even permutation, the asymptotic probability of an $n$-bit arbitrary permutation having $k$ free blocks is $1/(k! e)$. (It does not depend on $n$.)}
It is then our task to find a transformation to maximize the number.
For example, assume there is a map applied to a permutation matrix in Eq.\,(\ref{eq:mat-example}) as follows:
\begin{align}\label{eq:block-example}
\small
\setstretch{0.9}
\left(
{
	\arraycolsep=3pt
	\begin{array}{cccccccc}
	0 & 0 & 1 & 0 & 0 & 0 & 0 & 0 \\
	0 & 0 & 0 & 1 & 0 & 0 & 0 & 0 \\
	0 & 1 & 0 & 0 & 0 & 0 & 0 & 0 \\
	0 & 0 & 0 & 0 & 0 & 1 & 0 & 0 \\
	0 & 0 & 0 & 0 & 0 & 0 & 0 & 1 \\
	0 & 0 & 0 & 0 & 1 & 0 & 0 & 0 \\
	0 & 0 & 0 & 0 & 0 & 0 & 1 & 0 \\
	1 & 0 & 0 & 0 & 0 & 0 & 0 & 0
	\end{array}
}
\right)
\mapsto
\left(
{
	\arraycolsep=3pt
	\begin{array}{cc:cc:cc:cc}
	0 & 0 & 0 & 0 & 0 & 0 & \cellcolor[rgb]{0.7,0.7,0.7}0 & \cellcolor[rgb]{0.7,0.7,0.7}1 \\
	0 & 0 & 0 & 0 & 0 & 0 & \cellcolor[rgb]{0.7,0.7,0.7}1 & \cellcolor[rgb]{0.7,0.7,0.7}0 \\ \hdashline
	\cellcolor[rgb]{0.7,0.7,0.7}1 & \cellcolor[rgb]{0.7,0.7,0.7}0 & 0 & 0 & 0 & 0 & 0 & 0 \\
	\cellcolor[rgb]{0.7,0.7,0.7}0 & \cellcolor[rgb]{0.7,0.7,0.7}1 & 0 & 0 & 0 & 0 & 0 & 0 \\ \hdashline
	0 & 0 & 0 & 0 & \cellcolor[rgb]{0.7,0.7,0.7}1 & \cellcolor[rgb]{0.7,0.7,0.7}0 & 0 & 0 \\
	0 & 0 & 0 & 0 & \cellcolor[rgb]{0.7,0.7,0.7}0 & \cellcolor[rgb]{0.7,0.7,0.7}1 & 0 & 0 \\ \hdashline
	0 & 0 & \cellcolor[rgb]{0.7,0.7,0.7}0 & \cellcolor[rgb]{0.7,0.7,0.7}1 & 0 & 0 & 0 & 0 \\
	0 & 0 & \cellcolor[rgb]{0.7,0.7,0.7}1 & \cellcolor[rgb]{0.7,0.7,0.7}0 & 0 & 0 & 0 & 0
	\end{array}
}
\right).
\end{align}
The $2\times2$ substructures in gray color are called blocks.
Even (odd) blocks are $2\times2$ identity (off-diagonal) structures.
In one-line notation of the resulting matrix $(2,3,7,6,4,5,1,0)$, pairwise row numbers 2, 3 and 4, 5 are even blocks and 7, 6 and 1, 0 are odd blocks.\footnote{Note that the resulting matrix in Eq.\,(\ref{eq:block-example}) is not a tensor product of two smaller tensors, yet.
	Further application of $C\!X_{23}$ achieves four even blocks thereby completing the procedure.}


These pairs of row numbers play an important role hereafter, but there is a chance the notions confuse readers.
To avoid confusion, let us describe a way to construct an even block `$2,3$' beginning from $(7,2,0,1,5,3,6,4)$.
Using the $r_i$ notation, initially we have the pair $r_1=2$ and $r_5=3$.
First, we want to put 2 in the 0th column.
Applying $X_1 C\!X_{13} X_1$ leads to $(2,7,1,0,5,3,6,4)$.
This procedure can be interpreted as moving row number 2 from column position 1 to 0; $r_1 \mapsto r_0$.
Similarly, applying $C\!X_{31}$ then results in $(2,3,1,4,5,7,6,0)$; $r_5 \mapsto r_1$.
As already explained, applying gates can be interpreted as exchanging column numbers for fixed row numbers.
All the algorithms and subroutines introduced below have similar descriptions that first targeting a certain pair of row numbers, and then changing their column positions.
At this point, let us formally define such pairs.

\begin{definition}
	\label{def:rel-pair}
	A pair of row numbers $2j, 2j+1$ for $0 \le j \le 2^{n-1}-1$ is called a relevant row pair, denoted by $\langle 2j, 2j+1 \rangle$.
\end{definition}
%
Row numbers comprising a relevant row pair are called relevant row numbers.
Constructing a block is thus putting relevant row numbers appropriately side-by-side that are originally located apart.

In addition, we further define the following:

\begin{definition}
	\label{def:numbers}
	For a relevant row pair $\langle r_i,r_j \rangle$, the row numbers $r_i$ and $r_j$ are said to be occupied at
	\begin{itemize}[noitemsep]
		\item normal positions if $r_{i} \equiv i$ and $r_{j} \equiv j \mod 2$.
		\item inverted positions if $r_{i} \not\equiv i$ and $r_{j} \not\equiv j \mod 2$.
		\item interrupting positions otherwise.
	\end{itemize}
\end{definition}
To better understand the definitions, consider a permutation $(\colorbox[rgb]{0.7,0.7,0.7}{\makebox(5,6){$7$}}, \fbox{$2$}, \colorbox[rgb]{0.7,0.7,0.7}{\makebox(5,6){$0$}}, \fbox{$1$}, \colorbox[rgb]{0.7,0.7,0.7}{\makebox(5,6){$5$}}, \fbox{$3$}, \colorbox[rgb]{0.7,0.7,0.7}{\makebox(5,6){$6$}}, \fbox{$4$})$ and examine the relevant row pairs.
Relevant row numbers $r_2=0$ and $r_3=1$ are at normal positions as $2\equiv_2 0$ and $3\equiv_2 1$.
Another relevant numbers $r_7=4$ and $r_4=5$ are at inverted positions as $7 \not\equiv_2 4$ and $4 \not\equiv_2 5$.
Other numbers are at interrupting positions.
In simple terms, 0 and 1 are at normal positions as the small one is in the gray box, 4 and 5 are at inverted positions as the small one is in the white box, and other numbers are at interrupting positions as the numbers are in the same colored boxes.

It may seem plausible that the relevant numbers in normal (inverted) positions in the beginning likely end up in even (odd) blocks in the series of transformations for maximizing the number of blocks.
Indeed if we are careful enough, all the row numbers in the normal (inverted) positions in the beginning form even (odd) blocks at the end.
A formal description of the observation is as follows:

\begin{remark}
	\label{rem:maintain}
	When $C^m\!X$ gate is applied, the number of normal, inverted, and interrupting positions is conserved unless the gate is targeting the $n$-th bit.\footnote{More specifically, each row number remains still in its respective position unless the specified gates are involved, but we only need the fact that the number of such positions is conserved.}
\end{remark}
For example, a permutation $(7, 2, 0, 1, 5, 3, 6, 4)$ has four row numbers at interrupting positions; 2, 3 and 7, 6.
The action of $C\!X_{32}$ that does not \emph{target} the 3rd bit leads to the permutation $(7, 1, 0, 2, 5, 4, 6, 3)$, which still has four interrupting positions.
On the other hand, if $C\!X_{13}$ is applied, the resulting permutation $(7,2,0,1,3,5,4,6)$ will no longer involve any interrupting position, but instead, the number of normal and inverted positions will be increased by two each.
Details will not be covered, but we would point out that handling the interrupting positions is typically more expensive than dealing with normal or inverted positions.
Therefore in the next section, a preprocess that removes interrupting positions before getting into the main process will be introduced.
A way to deal with the inverted positions will also be introduced in the next section.
For now, let us restrict our attention to permutations that only have normal positions.
It will soon turn out that reversible gates targeting the last bit are not required at all in this section.

A natural way to design an algorithm for the size reduction could be to build blocks one-by-one, without losing already built ones.
Recall that in one-line notation for a given permutation $(\fbox{$r_0, r_1$}, \fbox{$r_2, r_3$}, \fbox{$r_4, r_5$}, \fbox{$r_6, r_7$}, \cdots)$, a block is a pair of row numbers $r_i$ that differ by 1 (with the odd one being larger) and both reside in one of the designated column positions (boxes).
Applying reversible gate swaps positions of at least two row numbers as in Eq.\,(\ref{eq:gate-actions}).
What we want to do is to construct a new block while maintaining already constructed ones, which very much resembles solving Rubik's Cube.
Each rotation in Rubik's Cube corresponds to the application of a logic gate.
The difference is that while the number of rotations is to be minimized typically in Rubik's Cube, we try to minimize the number of Toffoli gates.

\subsection{Glossary}\label{sec:glossary}
All necessary definitions and notations have been introduced, which we summarize below.
\begin{center}
	\setstretch{0.75}
	\begin{tabular}{ >{\raggedright}p{4.2cm}  >{\raggedright}p{5.5cm} }
		\hline
		$X_{i}$, $C\!X_{ij}$, $C\!X_{ijk}$	& NCT gates (Section\;\ref{sec:2-2})\tabularnewline
		$C\!X_{\mathcal{I}:k}$					& MCT gates (Section\;\ref{sec:2-1})\tabularnewline
		$R$           					& An ordered set of gates		\tabularnewline
		$P_{n}$           				& An $n$-bit substitution map	\tabularnewline
		$(r_{0}, r_1, \ldots, r_{2^{n}-1})$	& One-line notation of $P_{n}$	\tabularnewline
		Row number $r_{i}$			& Below Eq.\;(\ref{eq:gate-actions})	\tabularnewline
		Column number $i$ (of $r_{i}$) 	& Below Eq.\;(\ref{eq:gate-actions})	\tabularnewline
		Block							& Definition~\ref{def:block}\tabularnewline
		Block-wise position				& Definition~\ref{def:block}\tabularnewline
		Relevant row pair	        	& Definition\;\ref{def:rel-pair}	\tabularnewline
		Relevant row numbers            & Below Definition\;\ref{def:rel-pair} \tabularnewline
		Normal position	              	& Definition~\ref{def:numbers}	\tabularnewline
		Inverted position	         	& Definition~\ref{def:numbers}	\tabularnewline
		Interrupting position	        & Definition~\ref{def:numbers}	\tabularnewline
		$\v x$							& Binary string	of an integer $x$ \tabularnewline
		$x_{i}$                       	& The $i$-th bit of $\v x$ for $1 \leq i \leq n$ \tabularnewline
		Quality                       	& In Section\;\ref{sec:2-1} \tabularnewline
		\hline
	\end{tabular}
\end{center}

\subsection{Size Reduction Algorithm}\label{sec:3-4}
In this section we restrict our attention to a procedure $\mathcal{A}'_{\rm red}: \mathcal{P}'_n \rightarrow \{P_n: P_n = P_{n-1} \otimes I_2\} \times \mathcal{R}$, where $\mathcal{P}'_n$ is a set of all $n$-bit permutations of which row numbers are only in normal positions and $\mathcal{R}$ is a set of all ordered set of MCT gates.
The algorithm $\mathcal{A}_{\rm red}'$ is designed to obey three rules.
\begin{itemize}[noitemsep]
	\item A block is constructed one-by-one.
	\item The constructed block is allocated to the left in one-line notation.
	\item The number of left-allocated blocks should not decrease upon the action of any gate.
\end{itemize}
The meaning of the construction and the allocation is given in an example.
Assume a three-bit permutation $P_3 = (0, 1, 6, 3, 2, 5, 4, 7)$ is at hand.
The first block is already there, and we target the next block $(4, 5)$.
Applying $C\!X_{312}$ results in $P_3 \cdot C\!X_{312} = (0, 1, 6, 3, 2, 7, \colorbox[rgb]{0.7,0.7,0.7}{\makebox(11,6){$4,5$}})$.
One can see that a new block is \emph{constructed} in the 6th and the 7th columns at the cost of one Toffoli gate.
The constructed block is then \emph{allocated} in the 2nd and the 3rd columns by applying $C\!X_{21}$, i.e., $P_3 \cdot C\!X_{312} \cdot C\!X_{21} = (0, 1, \colorbox[rgb]{0.7,0.7,0.7}{\makebox(11,6){$4,5$}}, 2, 7, 6, 3)$.
In this example, only the construction costs a Toffoli gate whereas the allocation does not, but in general allocation also costs Toffoli gates.
In spite of the seemingly unnecessary costs of the allocation, we conclude that the left-allocation is more beneficial than leaving a block where it is constructed.
Detailed reasonings for the left-allocation will not be covered\,\footnote{Roughly explained, if the blocks are located randomly across the possible positions, it will get more and more difficult to construct a new block while maintaining the already constructed ones since cheaper logic gates tend to stir many positions. One way to mitigate the difficulty is to make blocks share the same bit value in the binary string of the column numbers so that certain controlled-gates leave such blocks unaffected by using that bit as a control.}.

In describing algorithms and subroutines, operations that update the permutation or the gate sequence will be frequently involved.
For a permutation $P_n$, an ordered set of gates $R= (a_1 ,\ldots, a_{|R|})$, and another ordered set of gates $S = (b_1 ,\ldots, b_{|S|})$, define a symbol $\cdot$ such that
\begin{align*}
(P_n,R)\cdot S &= \left(P_n \cdot b_1 \cdot b_2 \cdot \ldots \cdot b_{|S|}, (R;S)\right), \nn
(R;S) &= (a_1, \ldots, a_{|R|}, b_1, \ldots, b_{|S|}).
\end{align*}

\begin{figure}[H]
	\centering
	\includegraphics[width=0.8\textwidth]{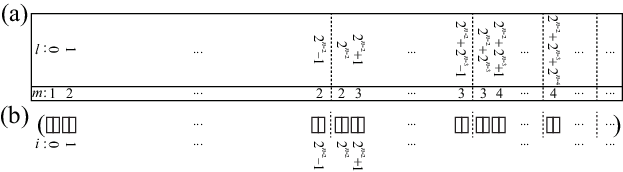}
	\caption{(a) Relation between $m$ and $l$. (b) Column positions (rectangle) and block-wise positions (two conjoined rectangles) in a permutation.
		Constructed blocks are to be allocated to the left, occupying block-wise positions denoted by $i$.
		When one tries to construct and allocate a new block at $i$-th position, $l$ equals $i$ and thus $m$ is determined as specified.
		Dashed vertical lines divide the number of column positions to be as ratios 1:1, 3:1, 7:1, and so on.}
	\label{fig:k-example}
\end{figure}

High-level description of the size reduction algorithm is as follows:
\begin{algorithm}[H]
	\small
	\setstretch{0.9}
	\caption{$\mathcal{A}_{\rm red}'$}
	\textbf{Input} $P_{n}$  \Comment{$P_n \in \mathcal{P}'_n$} \\
	\textbf{Output} $(P,R)$ \Comment{$P = P_{n-1} \otimes I_{2}$}\\
	\vspace{-8pt}
	\begin{algorithmic}[1]
		\State{$R \leftarrow \left(~\right)$} \Comment{An empty ordered set}
		\State{$P \leftarrow P_{n}$}
		\For{$i$ from $0$ to $2^{n-1} - 1$}                   
		\State{$\langle a, b\rangle \leftarrow$ $\mathsf{PICK}$$(P, i )$} \Comment{Pick a relevant row pair}
		\State{$\textit{S}_{C} \leftarrow$ $\mathsf{CONS}$($P$, $i$, $\langle a, b\rangle$)}
		\State{$(P,R) \leftarrow (P,R)\cdot S_C$}
		\State{$\textit{S}_{A} \leftarrow$ $\mathsf{ALLOC}$($P$, $i$, $a$)}
		\State{$(P,R) \leftarrow (P,R)\cdot S_A$}
		\EndFor
		\State{\textbf{return} $(P,R)$}
	\end{algorithmic}
	\label{alg:normal-decom}	
\end{algorithm}
Three subroutines $\mathsf{PICK}$, $\mathsf{CONS}$, and $\mathsf{ALLOC}$ will be examined in detail in the next subsection, and the working example is given in Appendix\;I, but it will be helpful to have some intuition behind the details here.
Define $m$ as the smallest positive integer satisfying
\begin{align}\label{eq:m}
	2l \le \sum_{j=1}^{m-1} 2^{n-j},
\end{align}
where $l$ is the number of left-allocated blocks and we define $m=1$ for $l=0$.
A pattern of how $m$ is determined depending on $l$ is illustrated in Fig.\;\ref{fig:k-example}.
Now notice that a group of left-allocated blocks distinguished by dashed lines in the figure shares the same bit values in their column numbers as $n$-bit binary strings.
For example, from $i=0$ to $2^{n-2}-1$ in the figure, the first bit of the column numbers is zero.
From $i=2^{n-2}$ to $2^{n-2}+2^{n-3}-1$, the first bit is one and the second bit is zero.
The point is, in constructing a new block, the (already) left-allocated blocks have some common properties.
The properties can be exploited to build a block without too much cost being paid, which we shall prove in Section\;\ref{sec:3-5} to be upper bounded by one $C^m\!X$ and a few $C\!X$ and $X$ gates.

In constructing a new block in the $i$-th iteration where $i$ is the iterator in Algorithm\;\ref{alg:normal-decom}, not all remaining relevant row pairs can be constructed as a block meeting the bound.
However, it can be shown that at least one pair that meets the bound always exists for all $i$.
Subroutine $\mathsf{PICK}$ takes care of passing an appropriate pair to $\mathsf{CONS}$ and $\mathsf{ALLOC}$ such that the quality bound can be met.

Allocation is similar to construction.
Let $\oplus$ denote the bit-wise XOR of binary strings.
Suppose we are to construct a block by conjoining two relevant row numbers of which column numbers are $\alpha$ and $\beta$, respectively, where $\v \alpha \oplus \v \beta \in \{ 0,1 \}^n$.
Construction can be understood as a process of changing $\alpha$ and $\beta$ so that $\v \alpha \oplus \v \beta = 00\cdots01$, without breaking left-allocated blocks.
By breaking, we mean some of the previously left-allocated blocks are no longer left-allocated, or the number of left-allocated blocks decreases.
Now, suppose we successfully constructed a block with the aforementioned relevant row pair, and we want to allocate it to the $i$-th block-wise position.
Let $c_i$ denote one of the column numbers at $i$-th block-wise position (regular squares in Fig.\;\ref{fig:k-example}).
Allocation is a process of changing $\alpha$ so that $\v c_{i} \oplus \v \alpha = 00\cdots0\!* ~(*\in \{0,1\})$, without breaking left-allocated blocks and with preserving $\v \alpha \oplus \v \beta = 00 \cdots 01$.
Therefore $\mathsf{ALLOC}$ is more or less the same as $\mathsf{CONS}$, and it will be shown in detail that in allocating a constructed block to $i$-th block-wise position, the cost is upper bounded by $C^{s(i)}\!X$ and a few $C\!X$ and $X$ gates, where $s(i)$ is a function such that $s(i) \le HW(\v i)$, where $HW(\v i)$ is the hamming weight of $\v i$.

How $\mathsf{PICK}$, $\mathsf{CONS}$, and $\mathsf{ALLOC}$ are designed so that the number of Toffoli gates involved is bounded is given in Section\;\ref{sec:3-5} as proofs.
The subroutine $\mathsf{PICK}$ is described below.
The role of $\mathsf{PICK}$ is to output a relevant row pair that requires the bounded number of Toffoli gates for the construction and the allocation.
\begin{algorithm}[H]
	\small
	\setstretch{0.9}
	\caption*{$\textbf{Subroutine 1}$ $\mathsf{PICK}$}
	\textbf{Input} $P_{n}$, $i$\\
	\textbf{Output} $\langle a,b \rangle$ \\
	\vspace{-8pt}
	\begin{algorithmic}[1]
		\State{$m \leftarrow$ $\mathsf{FINDM}$($i$) }
		\State{$k \leftarrow 2^n - 2^{n-m+1}$}
		\For{$j$ from $k$ to $2^n -2$}
		\State{$a \leftarrow r_j$}
		\State{$b \leftarrow \v a \oplus \v 1$} \Comment{$\v 1 = 00\ldots01$}
		\For{$t$ from $j+1$ to $2^n -1$}
		\State{$c \leftarrow r_t$}
		\State{\textbf{if} $b = c$ \textbf{then} \textbf{return} {$\langle a,b \rangle$}}
		\EndFor
		\EndFor
	\end{algorithmic}
	\label{alg:PICK}
\end{algorithm}
\noindent Termination of the subroutine will be shown to be guaranteed in the next subsection.
In line 1, $\mathsf{FINDM}$ computes $m$ from $i$ by letting $l=i$ in Eq.\;(\ref{eq:m}).
Next, $\mathsf{CONS}$ is described below.
It takes an input pair from $\mathsf{PICK}$, and then constructs a block.
It is designed following the proof of Lemma\;\ref{lem:construction}, in a way that the number of Toffoli gates involved is bounded.

\begin{algorithm}[H]
	\small
	\setstretch{0.9}
	\caption*{$\textbf{Subroutine 2}$ $\mathsf{CONS}$}
	\textbf{Input} $P_{n}, i, \langle a, b \rangle$ \\
	\textbf{Output} $S_C$ \\
	\vspace{-8pt}
	\begin{algorithmic}[1]
		\State{$S_C \leftarrow (~)$} \Comment{An empty ordered set}
		\State{$m \leftarrow$ $\mathsf{FINDM}$($i$)}
		\State{$(\alpha,\beta) \leftarrow$ $\mathsf{COL}$($P_n, a,b$)} \Comment{Find\! column\! numbers}
		\State{$\v \gamma \leftarrow \v \alpha \oplus \v \beta $}
		\State{$\delta \leftarrow 0$}
		\For{$j$ from $1$ to $n$}
		\State {\textbf{if} $\gamma_{j} = 1$ \textbf{then} $\delta \leftarrow j$; \textbf{break}}
		\EndFor
		\State{\textbf{if} $\delta = n$ \textbf{then} \textbf{return} $S_C$}
		\Comment{Already a block}
		\State{\textbf{if} $i_{\delta} = 1$ \textbf{then} $S_C \leftarrow S_C;X_{\delta}$}
		\Comment{Append $X_\delta$ to $S_C$}
		\For{$j$ from $\delta+1$ to $n-1$}
		\State{\textbf{if} $\gamma_{j} = 1$ \textbf{then} $S_C \leftarrow S_C;C\!X_{\delta j}$}
		\EndFor
		\State{\textbf{if} $i_{\delta} = 1$ \textbf{then} $S_C \leftarrow S_C;X_{\delta}$}
		\State{$\mathcal{I} \leftarrow \{n\}$}
		\For{$j$ from 1 to $m-1$}
		\State{$\mathcal{I} \leftarrow \mathcal{I} \cup \{j\}$}
		\EndFor
		\State{$S_C \leftarrow S_C;C\!X_{\mathcal{I}:\delta}$}
		\State{\textbf{return} $S_C$}
	\end{algorithmic}
	\label{alg:cons}
\end{algorithm}
%
\noindent In line 3, $\mathsf{COL}$ outputs column numbers $\alpha, \beta$ corresponding to the relevant row numbers $a, b$.
This procedure is necessary since what $\mathsf{CONS}$ essentially does is to swap columns as mentioned earlier.
%
Next, $\mathsf{ALLOC}$ is similarly described, which allocates the constructed block to the left.
It is designed following the proof of Lemma\;\ref{lem:allocation}, in a way that the number of Toffoli gates involved is bounded.

\begin{algorithm}[H]
	\small
	\setstretch{0.9}
	\caption*{$\textbf{Subroutine 3}$ $\mathsf{ALLOC}$}
	\textbf{Input} $P_{n}, i, a$ \\
	\textbf{Output} $S_{A}$\\
	\vspace{-8pt}
	\begin{algorithmic}[1]
		\State{$S_{A} \leftarrow (~)$}
		\State{$m \leftarrow$ $\mathsf{FINDM}$($i$)}
		\State{$\alpha \leftarrow$ $\mathsf{COL}$($P_n, a$)}
		\State{$\v \gamma \leftarrow \v \alpha \oplus \v i$}
		\State{$\delta \leftarrow 0$}
		\For{$j$ from $1$ to $n$}
		\State{\textbf{if} $\gamma_{j} = 1$ \textbf{then} $\delta \leftarrow j$; \textbf{break}}
		\EndFor
		\State{\textbf{if} $\delta = 0$ \textbf{then} \textbf{return} $S_{A}$}
		\Comment{Already allocated}
		\State{$\mathcal{I} \leftarrow \{~\}$}
		\For{$j$ from $\delta + 1$ to $n-1$}
		\State{\textbf{if} $\gamma_{j} = 1$ \textbf{then} $S_{A} \leftarrow S_{A};C\!X_{\delta j}$}
		\State{\textbf{if} $i_{j} = 1$ \textbf{then} $\mathcal{I} \leftarrow \mathcal{I} \cup \{j\}$}
		\EndFor
		\State{$S_{A} \leftarrow S_{A};C\!X_{\mathcal{I}:\delta}$}
		\State{\textbf{return} $S_{A}$}
	\end{algorithmic}
	\label{alg:alloc}
\end{algorithm}

\subsection{Complexity}\label{sec:3-5}
It has been sketched in Section\;\ref{sec:3-4} how the size reduction algorithm works.
Here we evaluate the time complexity and the quality of the output by introducing two lemmas, one for the construction and the other for the allocation.

\begin{proposition}\label{prop:pick}
	Let $l$ be the number of left-allocated blocks, and $m$ be the smallest positive integer satisfying Eq.\;(\ref{eq:m}).
	Define a function $h_n: \mathbb{N} \rightarrow \mathbb{Z}$, $h_n (x) = 2^{n} - 2^{n - (x-1)}$. There exists at least one relevant row pair $r_\alpha, r_\beta$ such that
	\begin{align}\label{eq:PICK}
	h_n(m) < \min(\alpha,\beta),
	\end{align}
	where $\alpha,\beta$ are column numbers of $r_\alpha, r_\beta$, respectively.
\end{proposition}
\begin{proof}
	It is nothing more than Pigeonhole principle.
	It is trivial for $m = 1$.
	For $m > 1$, there exist $2^n - 2l$ row numbers that do not form left-allocated blocks yet.
	Let us call them the remaining row numbers (row pairs).
	Assume there is no remaining row pair satisfying Eq.\;(\ref{eq:PICK}).
	Then at least one relevant row number in every remaining pair has to sit in column numbers no greater than $h_n(m)$.
	However, since $2l$ column numbers are already occupied by left-allocated blocks, there only exist $h_n(m) - 2l$ columns available for the row numbers to sit in.
	If we subtract $h_n(m) - 2l$ from half the number of remaining row numbers $(2^n - 2l)/2$,
	\begin{align*}
	\frac{(2^n - 2l)}{2} -\left(h_n(m) - 2l\right) = l - \frac{h_n(m-1)}{2}
	 > 0,
	\end{align*}
	where the inequality follows from the fact that $h_n(m-1)<2l \le h_n(m)$ by the definition of $m$.
	Therefore the assumption cannot be true and there must exist at least one pair that satisfies Eq.\;(\ref{eq:PICK}).
\end{proof}

\begin{lemma}
	\label{lem:construction}
	Let $i$ be an iterator used in Algorithm\;\ref{alg:normal-decom}, $l$ be the number of left-allocated blocks, and $m$ be the smallest positive integer satisfying Eq.\;(\ref{eq:m}). For a given permutation, a new block can be constructed by using at most $n-m-1$ $C\!X$, two $X$, and one $C^{m}\!X$ gates, without breaking left-allocated blocks.
\end{lemma}
\begin{proof}
	Let $\mathcal{C}=\{0,1,\cdots,2^{n-1} -1\}$ be a set of block-wise positions of an $n$-bit permutation, $\v i$ be the binary string of $i\in \mathcal{C}$, and $r_\alpha, r_\beta$ be relevant row numbers satisfying Eq.\;(\ref{eq:PICK}).
	Viewing their column numbers as binary strings $\v \alpha$ and $\v \beta$, the first $m-1$ bits of them are 1 (none if $m=1$),
	\begin{align}\label{eq:gamma-cons}
	\alpha_1 = \beta_1 = \alpha_2 = \beta_2 = \cdots = \alpha_{m-1}  = \beta_{m-1} = 1.
	\end{align}
	Let $\v \gamma = \v \alpha \oplus \v \beta$ and $\delta$ be the smallest integer such that $\gamma_\delta = 1$.
	It can be seen that $\delta>m-1$.
	Assume $\delta \ne n$, otherwise, the pair is already a block.
	Let $L_\delta$, $R_\delta$ be disjoint subsets of $\mathcal{C}$ such that $\mathcal{C} = L_\delta \bigcup R_\delta$ and the blocks residing in $L_\delta$ and $R_\delta$ are preserved under the actions of $C\!X_{\delta x}$ and $X_\delta C\!X_{\delta x} X_\delta$ gates for $x\in \{1,\ldots,\delta-1,\delta+1,\ldots,n \}$, respectively.
	A few $L_\delta$ and $R_\delta$ are illustrated in Fig.\;\ref{fig:delta}.
	\begin{figure}[t]
		\centering
		\includegraphics[width=0.65\textwidth]{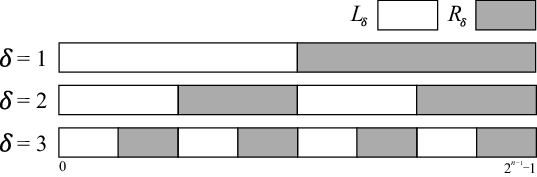}
		\caption{$L_\delta$ and $R_\delta$ for $\delta=1,2,3$. }
		\label{fig:delta}
	\end{figure}
	%
	
	At $i$-th iteration, note that $i$ (as a block-wise number) is either included in $L_\delta$ or $R_\delta$.
	If $i\in L_\delta$, an action of $C\!X_{\delta x}$, $\delta < x < n$ gate does not break left-allocated blocks since the blocks in $L_\delta$ are not affected at all and the blocks in $R_\delta$ change their respective positions within $<i$.
	If on the other hand $i\in R_\delta$, similarly an action of $X_\delta C\!X_{\delta x} X_\delta$, $\delta < x < n$ gate does not break left-allocated blocks since the blocks in $R_\delta$ are conserved and the blocks in $L_\delta$ change their respective positions within $<i$.
	Because $i_\delta$ tells if it is included in $L_\delta$ or $R_\delta$, we are able to apply the aforementioned gates without breaking left-allocated blocks.
	
	First $\delta-1$ bits of $\v \gamma$ are zeros due to the definition of $\delta$.
	Since $C\!X_{\delta x}$ or $X_\delta C\!X_{\delta x} X_\delta$, $\delta < x < n$ gate can be applied without breaking the left-allocated blocks, it is always possible to achieve $\gamma_\delta = \gamma_n = 1$ and $\gamma_k = 0$ for all $k\not \in \{\delta,n\}$, without involving any Toffoli gate.
	Once it is achieved, by applying $C\!X_{\mathcal{I}:\delta}$, $\mathcal{I}=\{ 1,2,\cdots,m-1,n \}$, we have $\v \gamma = 00\cdots 01$, which is by definition a block.
	Blocks located left to $i$ are unaffected by $C\!X_{\mathcal{I}:\delta}$ gate since the binary strings of their positions have at least one zero among the first $m-1$ bits.
\end{proof}

\begin{lemma}\label{lem:allocation}
	Let $i$ be an iterator used in Algorithm\;\ref{alg:normal-decom}, $l$ be the number of left-allocated blocks, and $m$ be the smallest positive integer satisfying Eq.\;(\ref{eq:m}). For given permutation with a block constructed by $\mathsf{CONS}$, the block can be allocated to the $i$-th block-wise position by using at most $(n-m)$ $C\!X$ and one $C^{HW(\v i')}\!X$ gates, without breaking left-allocated blocks, where $i' = i-h_n(m-1)/2$ for $m>1$ and $i'=i$ otherwise.
\end{lemma}
\begin{proof}
	Let $\v j$ be a binary string of the (block-wise) position $j$ of the block constructed by $\mathsf{CONS}$.
	Assume $j \ne i$, otherwise, the block is already left-allocated.
	Let $\v \gamma = \v i \oplus \v j$ and $\delta$ be the smallest integer such that $\gamma_\delta = 1$.
	Due to the property of the column numbers we begin with in $\mathsf{CONS}$, i.e., Eq.\;(\ref{eq:PICK}), the first $m-2$ bits of $\v i$ and $\v j$ are 1 (none if $m\le 2$),
	\begin{align}\label{eq:gamma-one-bits}
	i_1 = j_1 = i_2 = j_2 = \cdots = i_{m-2}  = j_{m-2} = 1,
	\end{align}
	and thus $\delta > m-2$.
	Since $i<j$ and $\gamma_1 = \gamma_2 = \cdots = \gamma_{\delta-1} = 0$, it is guaranteed that $i_\delta = 0$ and $j_\delta = 1$.
	Similar to the technique used in Lemma\;\ref{lem:construction}, $\v j$ can be transformed such that $\gamma_\delta=1$ and $\gamma_k=0$ for all $k \not \in \{\delta\}$ by using $C\!X_{\delta x}$ gates for $\delta < x < n$.
	Note that unlike in Lemma\;\ref{lem:construction} where $\v \alpha$ and $\v \beta$ both can be transformed by gates, here only $\v j$ is allowed to change.
	Once it is done, by applying $C\!X_{\mathcal{I}:\delta}$, $\mathcal{I} =\{ x \,|\, x>\delta , \;i_x = 1 \}$, the left-allocation is completed.
	Note that blocks residing in $L_\sigma$, $\sigma \in \{ x \,|\, x < \delta , \;i_x = 1 \}$ are conserved by an action of a gate whose target is $\delta$-th bit, up to changes of block-wise positions within.
	In addition, by the definition of $L_\sigma$, $i$ cannot be included in any of $L_\sigma$.
	Therefore, $C\!X_{\mathcal{I}:\delta}$ can be applied without breaking the left-allocated blocks.
\end{proof}

Upper bounds on the number of Toffoli gates in the output circuit naturally follow from two lemmas.
Recall that we have adopted the conversion formula $C^m\!X : C^2\!X = 1 : 2 m -3 $.
By counting the worst-case number of Toffoli gates in $\mathsf{CONS}$ and $\mathsf{ALLOC}$ for all iterations, it can be summarized as follows:

\begin{theorem}
	\label{thm:quality}
	The number of Toffoli gates in the output $R$ of Algorithm~\ref{alg:normal-decom} is upper bounded by $\mathcal{N}_{c}(n) + \mathcal{N}_{a}(n)$ for $n\ge 3$, where
	\begin{align}\label{eq:theorem-sec-3}
	\mathcal{N}_{c}(n) &= \sum_{i=2}^{n-1} \left(2i-3\right) \cdot 2^{n-i}, \nn
	\mathcal{N}_{a}(n) &= \sum_{j=2}^{n-2} {\sum_{i=2}^{n-j} \left(2i-3\right) \cdot {n-j \choose i}},
	\end{align}
	where $\mathcal{N}_a(3) =0$.
\end{theorem}

The time complexity of Algorithm\;\ref{alg:normal-decom} can also be estimated by the lemmas.
\emph{Assuming all gates exchange $2^n$ column numbers} (Section\;\ref{sec:6} for more discussion), the time complexity of the algorithm simply reads the total number of gates times $2^n$.
The maximum number of gates applied in each $\mathsf{CONS}$ and $\mathsf{ALLOC}$ is $O(n)$, and the number of $\mathsf{CONS}$ and $\mathsf{ALLOC}$ called in Algorithm\;\ref{alg:normal-decom} is $2^{n-1}$ each, and thus the asymptotic time complexity reads $O(n 2^{2n} )$. 

\section{Algorithm}\label{sec:4}
Key ideas have been introduced in Section\;\ref{sec:3}.
The only downside of Algorithm\;\ref{alg:normal-decom} is its inability to decompose arbitrary permutations.
This section is therefore mostly devoted to lifting the assumption that an input permutation consists only of row numbers in normal positions.
Lifting the assumption costs an extra number of Toffoli gates which is bounded by
\begin{align}\label{eq:extra-cost}
5\cdot2^{n-4} + 2n - 5 + \sum_{i=2}^{n-3} \left(2i-3\right) \cdot {n-3 \choose i}.
\end{align}
%

Lifting the assumption takes several independent steps, which we shall separately deal with in each subsection.
Details on various formulas appearing in this section are more or less the same as in the previous section, and thus will mostly be dropped.

\subsection{Heuristic Mixing}\label{sec:4-1}
Throughout Section\;\ref{sec:4} we will frequently refer to the ratio of the number of normal, inverted, and interrupting positions as the ratio $x:y:z$ such that $x+y+z=1$.
For example, all permutations handled in Section\;\ref{sec:3-4} have the ratio $1:0:0$.

The goal of the first step, Heuristic Mixing, is to transform an arbitrary permutation into one with the ratio $x:y:0.5$, where $x$ and $y$ are arbitrary.
The reason for the target ratio is as follows.
In the next step, we will use an algorithm to make the output ratio $0.5:0.5:0$ so that the problem effectively becomes two smaller instances.
The algorithm however requires that for its input ratio $x:y:z$, $x$ and $y$ should not be larger than 0.5.
One way to meet the requirement is to make $z=0.5$, thus Heuristic Mixing is applied first.

The method is to apply $C\!X$ gates a few times until the desired ratio is (nearly) achieved.
The following assumption is based on an observation: unstructured or randomly chosen permutations have a ratio close to $x:y:0.5$.
\begin{assumption}\label{assm:mixing}
	There exists a composite gate consisting of at most four $C\!X$ and one $C^{n-1}\!X$ gates that transforms a permutation such that the resulting permutation exhibits the ratio $x:y:0.5$.
\end{assumption}
%

Here we count $X_{i} C\!X_{ij} X_{i}$, $j\ne i$ (so called negative controlled gates) as $C\!X$ gate, too.
Let us call a composite gate consisting of $t$ $C\!X$ gates a depth-$t$ composite.
For example, counting the meaningful composites, there exist no depth-0 composites, $2n-2$ depth-$1$ composites,\footnote{The number of interrupting positions can be varied by $C\!X$ gate if the gate's target is the $n$-th bit. There exist $2n-2$ such $C\!X$ gates.} at most $(2n-2)\cdot 4 {n \choose 2}$ depth-2 composites, and so on.
As an algorithm, we may apply depth-$t$ composites one by one from $t=0$ to $t=4$ until the ratio hits $x:y:0.5$ exactly, or until all the composites are exhausted.
When the latter happens, we choose a permutation among the tried ones that are closest to the desired ratio, and then apply one $C^{n-1}\!X$ gate to meet $x:y:0.5$.

\begin{figure}[t]
	\centering
	\includegraphics[width=0.6\textwidth]{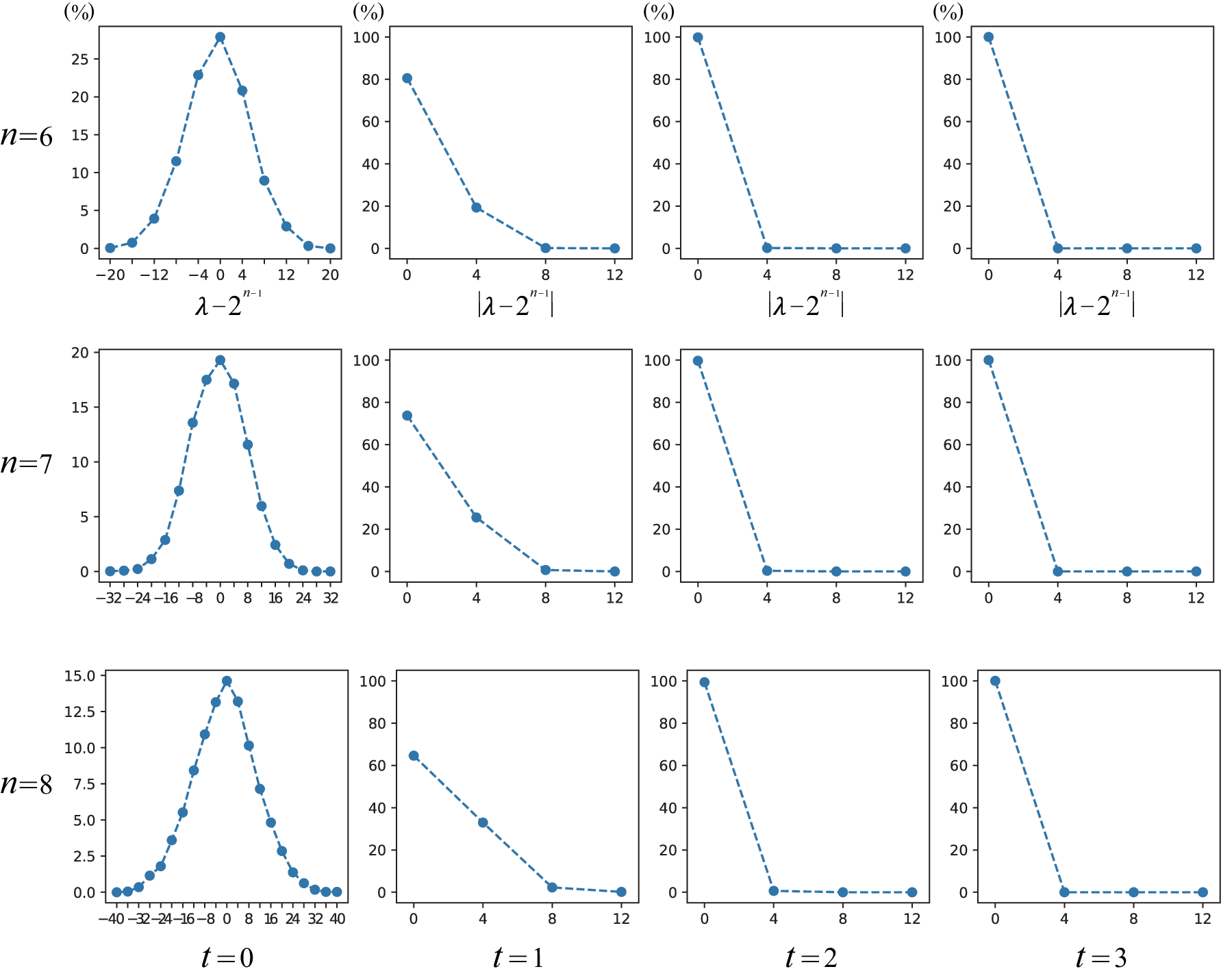}
	\caption{Sampling tests for $n\in\{6,7,8\}$ and $t\in\{0,1,2,3\}$. Each figure shows the percentage of samples as a function of $\lambda - 2^{n-1}$ (or $|\lambda - 2^{n-1}|$), where $\lambda$ is defined in the main text. In other words, figures show how far the permutations are from the ratio $x:y:0.5$ (and especially $\lambda - 2^{n-1} = 0$ means the ratio is exactly $x:y:0.5$). For $t=2$ where two CX gates have been applied, we were already able to find the desired permutations with high probability. The margin of error is $\pm1.55$\% at $95$\% confidence level. }
	\label{fig:mixing}
\end{figure}

We have carried out numerical tests on random samples for $t\in\{0,1,2,3\}$, plotting $\lambda - 2^{n-1}$ (or $|\lambda - 2^{n-1}|$), where $\lambda$ is the number of interrupting positions in a permutation we can get with depth-$t$ composites that is closest to the desired ratio.
As the results in Fig.\;\ref{fig:mixing} show, applying a few $C\!X$ composites likely leads to the desired ratio.

\subsection{Preprocessing}\label{sec:4-2}
The output of Heuristic Mixing is a permutation with the ratio $x:y:0.5$.
It is then further processed by an algorithm $\mathcal{A}_{\rm pre}$ to be the ratio $0.5:0.5:0$.

The idea is simple and we give an example first.
Consider the following $4$-bit permutation:
\begin{align*}
(3, \colorbox[rgb]{0.7,0.7,0.7}{\makebox(5,6){$10$}}, 14, 6,
\colorbox[rgb]{0.7,0.7,0.7}{\makebox(5,6){$12$}},2,\colorbox[rgb]{0.7,0.7,0.7}{\makebox(2,6){$0$}},15,
5,\colorbox[rgb]{0.7,0.7,0.7}{\makebox(2,6){$8$}},\colorbox[rgb]{0.7,0.7,0.7}{\makebox(5,6){$13$}},\colorbox[rgb]{0.7,0.7,0.7}{\makebox(2,6){$9$}},
\colorbox[rgb]{0.7,0.7,0.7}{\makebox(2,6){$1$}},4,7,\colorbox[rgb]{0.7,0.7,0.7}{\makebox(5,6){$11$}}),
\end{align*}
where eight interrupting positions are highlighted.
Move four of them to the left by a sequence of gates $C\!X_{13}$, $X_{1}$, $C\!X_{12}$, $C\!X_{13}$, $C\!X_{241}$, $C\!X_{32}$ in order, leading to
\begin{align*}
(\colorbox[rgb]{0.7,0.7,0.7}{\makebox(5,6){$13$}}, \colorbox[rgb]{0.7,0.7,0.7}{\makebox(2,6){$9$}}, \colorbox[rgb]{0.7,0.7,0.7}{\makebox(2,6){$1$}}, \colorbox[rgb]{0.7,0.7,0.7}{\makebox(5,6){$10$}},
7,6,5,\colorbox[rgb]{0.7,0.7,0.7}{\makebox(2,6){$8$}},
\colorbox[rgb]{0.7,0.7,0.7}{\makebox(2,6){$0$}},15,3,4,
14,\colorbox[rgb]{0.7,0.7,0.7}{\makebox(5,6){$11$}},\colorbox[rgb]{0.7,0.7,0.7}{\makebox(5,6){$12$}},2).
\end{align*}
Applying another sequence\! $X_{1}$, $X_{2}$, $C\!X_{124}$, $X_{1}$, $X_{2}$ results in
\begin{align*}
(9,13,10,1,7,6,5,8,0,15,3,4,14,11,12,2).
\end{align*}

Here we only sketch the mechanism, rather than elaborating on it.
Due to the way the interrupting position is defined, it can only be an even number; (hypothetical) change of a `single' row number leads to either $- 2, +0$, or $+2$ change to the number of interrupting positions.
Since at least two row numbers are exchanged by the logic gates, the number of interrupting positions can only vary by multiples of 4.
Note that an action of $C^{n-1}\!X$ gate that swaps two row numbers can change the number of interrupting positions by at most 4.
Considering along the line, an action of $C^{2}\!X$ gate that swaps $2^{n-2}$ row numbers can change the number by at most $2^{n-1}$.
Since the number of interrupting positions of an input permutation is $2^{n-1}$, at best a single Toffoli can achieve the desired ratio.
However, the best case hardly happens naturally, and we need to manipulate the input before applying the Toffoli gate.
It is complicated to give details, but we claim that a procedure similar to repeating $\mathsf{CONS}$ and $\mathsf{ALLOC}$ one-quarter times of that of $\mathcal{A}'_{\rm red}$ is enough.
In addition to its ability to eliminate interrupting positions, the procedure can also fine-tune the number of normal and inverted positions.
In fact, $\mathcal{A}_{\rm pre}$ can turn the ratio $x:y:0.5$ into $x\pm \Delta_x : y\pm \Delta_y : 0$ where $0 \le \Delta_x, \Delta_y \le 0.5$ as desired. 
In the following algorithm $\mathcal{A}_{\rm pre}$, we may simply set the output ratio to be 0.5:0.5:0.

The procedure is described as follows:
\begin{algorithm}[H]
	\small
	\setstretch{0.9}
	\caption{$\mathcal{A}_{\rm pre}$}
	\textbf{Input} $P_{n}$  \Comment{The ratio is $x:y:0.5$}\\
	\textbf{Output} $(P,R)$ \\
	\vspace{-8pt}
	\begin{algorithmic}[1]
		\State{$R \leftarrow \left(~\right)$}
		\State{$P \leftarrow P_{n}$}
		\For{$i$ from $0$ to $2^{n-3}-1$}
		\State{$\langle a, b\rangle \leftarrow$ $\mathsf{PRE\_PICK}$$(P, i)$}
		\State{$\textit{S}_{C} \leftarrow$ $\mathsf{CONS}$($P$, $i$, $\langle a, b \rangle$)}
		\State{$(P,R) \leftarrow (P,R)\cdot S_C$}
		\State{$\textit{S}_{A} \leftarrow$ $\mathsf{ALLOC}$($P$, $i$, $a$)}
		\State{$(P,R) \leftarrow (P,R)\cdot S_{A}$}
		\EndFor
		\State{$(P,R) \leftarrow (P,R) \cdot (X_{1} , X_{2} , C\!X_{12n} , X_{2} , X_{1})$}
		\State{\textbf{return} $(P,R)$}
	\end{algorithmic}
	\label{alg:preprocess}
\end{algorithm}
\noindent $\mathsf{PRE\_PICK}$ works in a similar way to $\mathsf{PICK}$, but the output is not a relevant row pair but a certain pair of row numbers at interrupting positions.
We will not cover the details on the pair, but with abuse of notation we let $\langle \cdot ,\cdot \rangle$ denote the pair, too.
Interested readers may refer to the implemented code\;\cite{code}.
$\mathsf{CONS}$ and $\mathsf{ALLOC}$ work exactly in the same way as in Section\;\ref{sec:3-4}.
The number of Toffoli gates involved in Algorithm\;\ref{alg:preprocess} is bounded by $3\cdot 2^{n-4} -1 + \sum_{i=2}^{n-3} \left(2i-3\right) \cdot {n-3 \choose i}$.

\subsection{Generalized Size Reduction Algorithm}\label{sec:4-3}
After the mixing and the preprocessing, the permutation has the ratio $0.5:0.5:0$.
Now the problem can be thought of as two subproblems. 
We will first construct and allocate only \emph{even blocks} out of row numbers that are in normal positions.
Since $\mathsf{CONS}$ and $\mathsf{ALLOC}$ do not require any controlled gate targeting the $n$-th bit, the number of normal or inverted positions is conserved by Remark\;\ref{rem:maintain}.
Once $2^{n-2}$ even blocks are left-allocated, the remaining row numbers that are only in inverted positions are constructed and allocated to be \emph{odd blocks}.
In the end, we would have $2^{n-2}$ even blocks on the left half and $2^{n-2}$ odd blocks on the right half.
Applying one $C\!X_{1n}$ gate completes the size reduction.

The size reduction algorithm is described as follows:
\begin{algorithm}[H]
	\small
	\setstretch{0.9}
	\caption{$\mathcal{A}_{\rm red}$}
	\label{alg:decom}
	\textbf{Input} $P_{n}$ \Comment{The ratio is $0.5:0.5:0$}	\\
	\textbf{Output} $(P,R)$
	\Comment{$P=P_{n-1} \otimes I_2$} \\
	\vspace{-8pt}
	\begin{algorithmic}[1]
		\State{$R \leftarrow \left(~\right)$}
		\State{$P \leftarrow P_{n}$}
		\For{$i$ from $0$ to $2^{n-2}-1$} \Comment{First Part}
		\State{$\langle a, b\rangle \leftarrow$ $\mathsf{N\_PICK}$$(P, i)$} \Comment{Pick a normal pair}
		\State{$\textit{S}_{C} \leftarrow$ $\mathsf{CONS}$($P$, $i$, $\langle a, b\rangle$)}
		\State{$(P,R) \leftarrow (P,R)\cdot S_C$}
		\State{$\textit{S}_{A} \leftarrow$ $\mathsf{ALLOC}$($P$, $i$, $a$)}
		\State{$(P,R) \leftarrow (P,R)\cdot S_A$}
		\EndFor
		\For{$i$ from $2^{n-2}$ to $2^{n-1}-1$} \Comment{Second Part}
		\State{$\langle a, b\rangle \leftarrow$ $\mathsf{PICK}$$(P, i)$}
		\State{$\textit{S}_{C} \leftarrow$ $\mathsf{CONS}$($P$, $i$, $\langle a, b\rangle$)}
		\State{$(P,R) \leftarrow (P,R)\cdot S_C$}
		\State{$\textit{S}_{A} \leftarrow$ $\mathsf{ALLOC}$($P$, $i$, $a$)}
		\State{$(P,R) \leftarrow (P,R)\cdot S_A$}
		\EndFor
		\State{$(P,R) \leftarrow (P,R) \cdot (C\!X_{1n})$}
		\State{\textbf{return} $(P,R)$}
	\end{algorithmic}
\end{algorithm}

The first part deals with row numbers in normal positions to construct and allocate even blocks only.
Therefore, $\mathsf{N\_PICK}$ has to output a relevant row pair in normal positions.
Accordingly, the working mechanism of $\mathsf{N\_PICK}$ is a bit different from that of $\mathsf{PICK}$\;\cite{code}.
Quality bounds given by Proposition\;\ref{prop:pick}, Lemma\;\ref{lem:construction}, and Lemma\;\ref{lem:allocation} are no longer valid in the first part, but overall it only adds $2^{n-3}$ to the quality bound of $\mathcal{A}'_{\rm red}$.
Details on $\mathsf{N\_PICK}$ and related quality bound are not discussed here.
The second part works exactly the same as Algorithm\;\ref{alg:normal-decom} except now the row numbers are only in inverted positions instead of normal positions.

The additional quality factor, $2^{n-3}$, can hardly be met in average instances, and in practice the quality difference between outputs of $\mathcal{A}'_{\rm red}$ and $\mathcal{A}_{\rm red}$ is negligible.
Note that the point of mixing and preprocessing is to transform an arbitrary permutation so that the result has an appropriate form for the decomposition.
If a given permutation is already well-suited for $\mathcal{A}'_{\rm red}$, or $\mathcal{A}_{\rm red}$, or something similar, mixing and preprocessing can be skipped.

\subsection{Algorithm for Synthesis}\label{sec:4-4}
Combining Heuristic Mixing, Preprocessing, and $\mathcal{A}_{\rm red}$, one can easily come up with an algorithm for reversible logic circuit synthesis as follows:
\begin{algorithm}[H]
	\small
	\setstretch{0.9}
	\caption{$\mathcal{A}_{\rm syn}$}
	\textbf{Input} $P_{n}$ \\
	\textbf{Output} $R$ \\
	\vspace{-8pt}
	\begin{algorithmic}[1]
		\State{$R \leftarrow \left(~\right)$}
		\For{$i$ from $n$ to 3}
		\State{$(P_{i}, R_{\rm mix}) \leftarrow \mathcal{A}_{\rm mix}(P_{i})$}\Comment{Heuristic Mixing}
		\State{$(P_{i}, R_{\rm pre}) \leftarrow \mathcal{A}_{\rm pre}(P_{i})$}
		\State{$(P_{i}, R_{\rm red}) \leftarrow \mathcal{A}_{\rm red}(P_{i})$}
		\State{$P_{i-1} \leftarrow $ $\mathsf{REDUCE}$$(P_{i})$}   \Comment{$P_{i} = P_{i-1} \otimes I_{2}$}
		\State{$R \leftarrow R$;$R_{\rm mix}$;$R_{\rm pre}$;$R_{\rm red}$} \Comment{Append all}
		\EndFor
		\State{$R \leftarrow R$;$\mathsf{SEARCH}$$(P_{2})$}
		\Comment{Exhaustive search for $P_{2}$}
		\State{\textbf{return} $R$}
	\end{algorithmic}
	\label{alg:synthesis}
\end{algorithm}
%

Let the quality bound given by Theorem\;\ref{thm:quality} be $A(n)$ and the cost Eq.\;(\ref{eq:extra-cost}) be $B(n)$ for an $n$-bit permutation.
The number of Toffoli gates required for the size reduction $P_n \mapsto P_{n-1}$ is upper bounded by
\begin{align}\label{eq:reduction-bound}
A(n)+B(n).
\end{align}
Therefore the quality bound on the output $R$ of Algorithm\;\ref{alg:synthesis} is given by $\sum_{x=n}^3 \left(A(x) + B(x) \right)$.

Time complexity of $\mathcal{A}_{\rm mix}$ can be estimated by counting the number of composites times $2^n$, which reads $O(n^7 2^n)$.
Preprocessing roughly does one-quarter of what the size reduction performs, thus the time complexity is $O(n 2^{2n-2} )$.
Time complexities of $\mathcal{A}'_{\rm red}$ and $\mathcal{A}_{\rm red}$ are the same.
In $\mathcal{A}_{\rm syn}$, the first size reduction step dominates the overall running time because each time the effective size gets smaller, time to transform the permutation becomes exponentially faster.
The time complexity of size reduction steps in $\mathcal{A}_{\rm syn}$ is dominated by $\mathcal{A}_{\rm red}$ with $O(n 2^{2n})$, thus the time complexity of $\mathcal{A}_{\rm syn}$ is $O(n 2^{2n})$.

%

\section{Benchmark and Application}\label{sec:5}
Algorithm\;\ref{alg:synthesis} can be considered as a very basic algorithm that makes use of the size reduction idea.
Indeed we have focused on simplifying the algorithm so that the design criteria are as transparently brought out as possible.
When it comes to optimizations we have noticed and indeed gone through a number of directions, for example exploiting partial search or statistical properties or undoing and retrying, but then what we confronted at the end was way too complicated descriptions for the algorithm.
Most optimization options we have considered earlier thus have been dropped at the end, not only from the paper but also from the source code so that the code can be a natural reflection of the descriptions given in Sections\;\ref{sec:3} and\;\ref{sec:4}.

Nevertheless, one optimization option is still implemented in our code.
Briefly explained, as shown in Subroutines $\mathsf{CONS}$ and $\mathsf{ALLOC}$, there is a possibility that a block is already existing at the iteration.
We call such cases free constructions or free allocations, or simply free blocks without distinction. 
To utilize free blocks, let us review how the algorithm works.

\begin{figure}[t]
	\centering
	\includegraphics[width=0.48\textwidth]{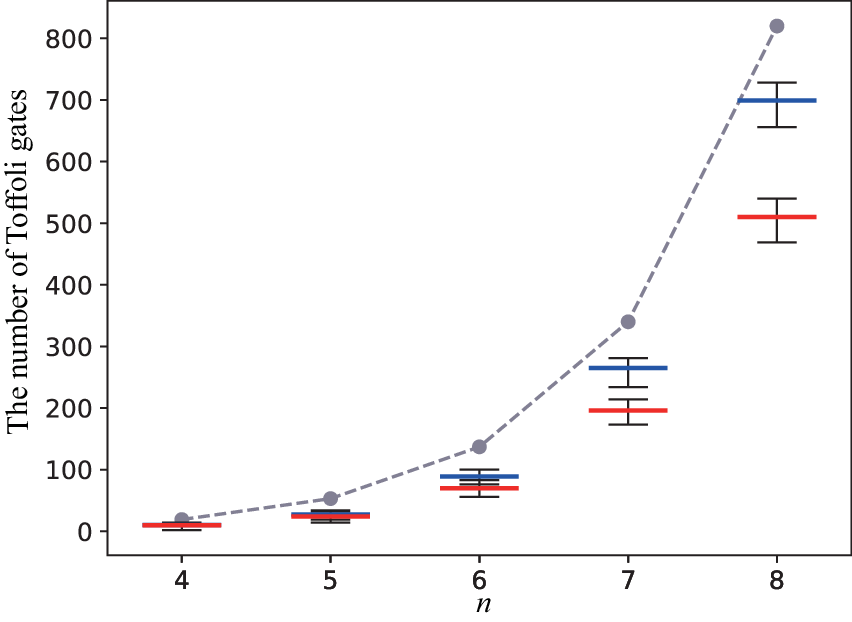}
	\caption{Quality bounds on size reduction of $n$-bit permutations (Eq.\;(\ref{eq:reduction-bound})) joined by dashed lines, and the statistical averages of the output qualities denoted by the red and blue bars for 1000 randomly generated instances for each $n$ with depth-0 (blue) and depth-1 (red) partial search. Min-Max values are also marked with black bars.}
	\label{fig:average}
\end{figure}

In $i$-th iteration in Algorithm\;\ref{alg:decom}, $\mathsf{N\_PICK}$ or $\mathsf{PICK}$ chooses one relevant row pair and it is processed to be a left-allocated block.
The resulting permutation may or may not have free blocks in $(i+1)$-th iteration.
Notice at this point that instead of specifying only one relevant row pair for the block, we may try all possible relevant row pairs and rank them based on certain criteria, for example by the number of free blocks upon the left-allocation of the candidate pair (block).\footnote{In fact, it is much more complicated than just counting the number of free blocks. Finding a sophisticated cost metric for evaluating free blocks is already a nontrivial task for optimization.}
Let us call it depth-$1$ partial search at $i$-th iteration.
Fig.\;\ref{fig:average} shows the number of Toffoli gates resulting from the size reduction of 1000 randomly generated permutations for $n=4,5,6,7,8$ employing depth-0 and depth-1 partial search for all blocks, together with the upper bound for the reference.
If computational resources allow, one may try searching for a better pair by considering free blocks upon the construction and allocation of blocks not only at the $i$-th block-wise position but also up to even more later blocks.

To be more specific, define $d(i)$ a search depth at position $i$.
For $d(i)=0$, the partial search option is off.
For $d(i)=1$, we do depth-1 partial search at $i$ as described earlier.
For $d(i)>1$, we try all the possible combinations of the blocks up to $i+d(i)-1$ block-wise position and pick the best one, and call it depth-$d(i)$ partial search at $i$.
For example, setting $d (i) =2$ for all $i\in\{0,1,\ldots, 2^{n-1}-2 \}$ is to carry out depth-2 partial search in constructing and allocating each block. 
The quality of the output circuit gets better as $d(i)$ gets larger, but the time complexity becomes worse.
Roughly estimated, for a constant $d(i) = d$ for all $i$, the time complexity reads $O\left(n 2^{(2+d)n} \right)$.

Since the number of remaining relevant row numbers decreases as the iteration goes on, it becomes even easier to do the search for larger search depth in later iterations.
%
Let $r(i)$ denote the number of remaining relevant row numbers in normal positions at $i$-th iteration.
Let $d_j$ be a search depth at $i$-th iteration such that $2^{j-1} < r(i) \le 2^{j}$.
In constructing and allocating even blocks, we may apply depth-$d_j$ partial search.
In the early stages, one should set small $d_j$ and let $d(i) = d_j$ so that the searching is tractable, but in later stages of left-allocating even blocks, $d_j$ can be larger.
Similar search depth can be defined for odd blocks, and we use the same symbol $d_j$ for either case.
In our code, we basically set $d(i)=d_j$ and let users control $d_j$ as external parameters for each $j \in \{1, \ldots, n\}$, except $d(i)$ for $i\ge 2^{n-1}-9$.\footnote{It is set such that the last few blocks are exhaustively searched, which has a nontrivial effect on the output quality considering the statistical properties and Lemma\;\ref{lem:construction}. Especially for even permutations, output circuits contain either zero or two $C^{n-1}\!X$ gates. The setting likely eliminates the chance of the latter taking place. About the statistical properties, for example, it can be shown that the last four blocks cost the \emph{minimal} number of Toffoli gates if no free block exists upon the construction and allocation of a block in $(2^{n-1}-5)$-th iteration. It seems many iteration stages can utilize such statistical properties.}
When $d_j$ is set as a constant $d$ for all $j$, we simply say depth-$d$ partial search is applied.

\begin{table*}[th]
	\centering
	\caption{Synthesis results for various structured and unstructured functions with depth-2 partial search. In subcolumns, \#in, \#out, \#grb read the number of input, output, and garbage bits, respectively. QC stands for Quantum Cost which is one of the widely accepted cost metrics\;\cite{RCViewer,rev-page2}, and \#TOF is the number of Toffoli gates. Most previous results have not reported \#TOF, some of which we were unable to retrieve ourselves. Ratio columns show the ratio of this work to the previous result in each respective metric. DES, Skipjack, and KHAZAD are S-box functions used in cryptography. For DES, the values are averaged since there are eight different S-boxes.}
	\label{tab:benchmarks}
	\linespread{1}
	\setlength\tabcolsep{2pt}
	\begin{tabular}{
			>{\centering}p{0.5cm} | >{\centering}p{2.0cm} |
			>{\centering}p{0.5cm} | >{\centering}p{0.7cm} | >{\centering}p{0.7cm} | >{\raggedleft}p{1.8cm}| >{\raggedleft}p{1.2cm} |
			>{\centering}p{0.5cm} | >{\centering}p{0.7cm} | >{\centering}p{0.7cm} | >{\raggedleft}p{1.1cm}| >{\raggedleft}p{1.2cm} |
			>{\raggedleft}p{0.9cm}| >{\raggedleft}p{1.2cm} }
		&\multirow{2}{*}{Functions}& \multicolumn{5}{c|}{Previous work}	& \multicolumn{5}{c|}{This work} &	\multicolumn{2}{c}{Ratio}\tabularnewline \cline{3-14}
		&& {\fontsize{8pt}{10pt} \selectfont \#in } & {\fontsize{8pt}{10pt} \selectfont \#out } & {\fontsize{8pt}{10pt} \selectfont \#grb } & \centering{QC}	& \centering{\#TOF} & {\fontsize{8pt}{10pt} \selectfont \#in } & {\fontsize{8pt}{10pt} \selectfont \#out } & {\fontsize{8pt}{10pt} \selectfont \#grb } & \centering{QC} & \centering{\#TOF} & \centering{QC} & \centering{\#TOF} \tabularnewline \hline
		\multirow{3}{*}{\rotatebox[origin=c]{90}{Unstructured~~~~~~~~~~~~}}
		&URF1			& 9	 & 9 & 0 & 17002\;\cite{SZM10} & 2225	& 9 & 9 & 0 &13318 & 2029 & 78\% & 91\% \tabularnewline
		&URF2			& 8	 & 8 & 0 & 7083\;\cite{SZM10}	 & 892	& 8 & 8 & 0 &5510  &  803 & 78\% & 90\% \tabularnewline
		&URF3			& 10 & 10& 0 & 37517\;\cite{SZM10} & 4997	& 10& 10& 0 &33541 & 4898 & 89\% & 98\% \tabularnewline
		&URF4			& 11 & 11& 0 & 160020\;\cite{Sae08}& 32004& 11& 11& 0 &77616 & 11706& 49\% & 37\% \tabularnewline
		&URF5			& 9	 & 9 & 0 & 14549\;\cite{SZM10} & 1964	& 9 & 9 & 0 &9863  & 1366 & 68\% & 70\% \tabularnewline
		&$n$thPrime7	& 7	 & 7 & 0 & 2284\;\cite{Zak16}	 & 334	& 7 & 7 & 0 &1887  &  281 & 83\% & 84\% \tabularnewline
		&$n$thPrime8	& 8	 & 8 & 0 & 6339\;\cite{Zak16}	 & 930	& 8 & 8 & 0 &5165  &  691 & 81\% & 74\% \tabularnewline
		&$n$thPrime9	& 10 & 9 & 1 & 17975\;\cite{SZM10} & 2392	& 9 & 9 & 0 &12189 & 1762 & 68\% & 74\% \tabularnewline
		&$n$thPrime10	& 11 & 10& 1 & 40299\;\cite{SZM10} & 5302	& 10& 10& 0 &28630 & 4003 & 71\% & 76\% \tabularnewline
		&$n$thPrime11	& 12 & 11& 1 & 95431\;\cite{SZM10} & 12579& 11& 11& 0 &63314 & 9269 & 66\% & 74\% \tabularnewline
		&DES	& - & - & - & - & - & 6 & 4 & 2 & 743.88 & 102.63 & - & - \tabularnewline
		&Skipjack	& - & - & - & - & - & 8 & 8 & 0 & 5562 & 791 & - & - \tabularnewline
		&KHAZAD	& - & - & - & - & - & 8 & 8 & 0 & 5411 & 794 & - & - \tabularnewline\hline\hline
		\multirow{3}{*}{\rotatebox[origin=c]{90}{Structured~~~~~~~~~~~~~~~~~~}}
		&HWB7	& 7 & 7 & 0 & \textit{poly(n)}\;\cite{HWB20} & 249	& 7 & 7 & 0 & 1979 & 288 & - & - \tabularnewline
		&HWB8	& 8 & 8 & 0 & \textit{poly(n)}\;\cite{HWB20}& 586 & 8 & 8 & 0 & 5439 & 788 & - & - \tabularnewline
		&HWB9	& 9 & 9 & 0 & \textit{poly(n)}\;\cite{HWB20}& 1853 & 9 & 9 & 0 & 13333 & 2010 & - & - \tabularnewline
		&HWB10	& 10 & 10 & 0 & \textit{poly(n)}\;\cite{HWB20}& 3411 & 10 & 10 & 0 & 33120 & 4885 & - & - \tabularnewline
		&HWB11	& 11 & 11 & 0 & \textit{poly(n)}\;\cite{HWB20}& 8536 & 11 & 11 & 0 & 75961 & 11497 & - & - \tabularnewline
		&aj-e	& 4 & 4 & 0 & 30\;\cite{rev-page} & 5 & 4 & 4 & 0 & 108 & 12 & 360\% & 240\% \tabularnewline
		&graycode6	& 6 & 6 & 0 & 5\;\cite{WilleSPD12}& 0 & 6 & 6 & 0 & 13 & 0 & 260\% & - \tabularnewline
		&ham3	& 3 & 3 & 0 & 9\;\cite{WilleSPD12}& 1 & 3 & 3 & 0 & 18 & 1 & 200\% & 100\% \tabularnewline
		&ham7	& 7 & 7 & 0 & 49\;\cite{MaslovDM07}& 6 & 7 & 7 & 0 & 325 & 22 & 663\% & 367\% \tabularnewline
		&mod5mils	& 5 & 5 & 0 & 13\;\cite{WilleSPD12}& 6\;\cite{DKRFM21} & 5 & 5 & 0 & 94 & 8 & 723\% & 133\% \tabularnewline
		&cycle10	& 12 & 12 & 0 & 1198\;\cite{miller04}& 98 & 12 & 12 & 0 & 78448 & 11905 & 6548\% & 130\% \tabularnewline
		&plus127mod$2^{13}$	& 13 & 13 & 0 & 35348\;\cite{MillerS12} & - & 13 & 13 & 0 & 13669 & 609 & 39\% & - \tabularnewline
		&plus63mod$2^{12}$ & 12 & 12 & 0 & 14652\;\cite{MillerS12} & - & 12 & 12 & 0 & 8771 & 536 & 60\% & - \tabularnewline
		&plus63mod$2^{13}$ & 13 & 13 & 0 & 19566\;\cite{MillerS12} & - & 13 & 13 & 0 & 18517 & 1177 & 95\% & - \tabularnewline \hline
	\end{tabular}
\end{table*}

Reversible logic circuits for known functions have been synthesized by the algorithm, which are summarized and compared with the previous results when available in Table\;\ref{tab:benchmarks}.
Depth-2 partial search option is applied to the functions.
Benchmark results with different $d_j$ options are to be posted on\;\cite{code}.\footnote{To get a sense of the running time, on a commercial laptop, eight-bit permutation URF2 takes 6s, 16s, 228s for the depth-0, 1, 2 partial search, respectively. Better runtime data can be found in\;\cite{code}.}
Notice that the algorithm works well only for the unstructured functions compared with the previous constructions.
About the hidden weighted bit (HWB) function, it has recently been shown by\;\cite{HWB20} that the circuit of size $O(n^{6.42})$ can be synthesized by using the reversible gates, or the circuit of size $O(n^{2})$ can be constructed by using quantum unitary gates.
We were unable to generate the concrete circuit of\;\cite{HWB20} ourselves for the comparison, but since their circuit is polynomial in $n$ in size, ours should be less efficient.

\begin{table*}[th]
	\caption{Comparison of different circuit designs for AES S-box given in\;\cite{aes20b} and a circuit obtained by the algorithm with depth-4 partial search. The third to the sixth columns read the number of CNOT gates, single qubit Clifford gates, T gates, and measurements. TD and W mean T-depth and the number of logical qubits, respectively. }
	\centering
	\setlength\tabcolsep{2.5pt}
	\begin{tabular}{>{\centering}p{1.8cm}| >{\centering}p{2.2cm}| >{\raggedleft}p{0.9cm}| >{\raggedleft}p{1.1cm}| >{\raggedleft}p{0.8cm}| >{\raggedleft}p{0.8cm}| >{\raggedleft}p{0.8cm}| >{\raggedleft}p{0.7cm}}
		\hline
		Source				& \centering{Type}	& \centering{\#CX}	& \centering{\#1qC}	& \centering{\#T}	& \centering{\#M}	& \centering{TD}	& \centering{W}		\tabularnewline \hline
		\;\cite{aes16}		& out-of-place	& 8683	& 1023	& 3854	& 0		& 217	& 44	\tabularnewline
		\;\cite{aes20}		& out-of-place	& 818	& 264	& 164	& 41	& 35	& 41	\tabularnewline
		\;\cite{aes20b}		& out-of-place	& 654	& 184	& 136	& 34	& 6		& 137	\tabularnewline
		This work	& in-place	& 12466	& 1787	& 5089	& 0		& 579	& 15	\tabularnewline \hline
	\end{tabular}
	\label{tab:aes-qsharp}
\end{table*}
The algorithm is applied to find reversible circuits for cryptographic S-boxes.
Among various S-boxes, one that has attracted the most attention from the research community is undoubtedly AES S-box\;\cite{aes16,aes20b,aes20,aes20c}.
Table 1 in Jaques et al.'s work\;\cite{aes20b} neatly summarizes a few notable circuit designs from the literature, and we add one more entry to it as in Table\;\ref{tab:aes-qsharp} with the same Q\# options being used as the first entry.
The design we have added in the table is first obtained by the algorithm proposed, and then it went through manual optimizations\;\cite{barenco95,he17} before Q\# estimates the resources.\footnote{The reason for the manual optimization is that the output circuit of the proposed algorithm consists of MCT gates, whereas all the other entries are written in terms of smaller gates such as T. Therefore, we first manually transform MCT gates into NCT gates, and then Q\# is applied.}
A detailed circuit can be found in\;\cite{code}.
Note that we could get an even smaller T-depth at the cost of increasing the number of measurements, but we post the one without any measurements.
Here we would point out that our design works \emph{in-place}\;\cite{draper06}, meaning that it does not need extra bits for writing the outputs.
In-place design of the S-box likely leads to a quantum oracle with a minimal number of qubits, but we leave the task of constructing an oracle as one of the future directions to explore.
Note however that in Grover's algorithm it is generally more important to optimize depth than width considering the quadratic speedup\;\cite{grover97,zalka00,aes18}.

Efficient reversible circuits can be found for S-boxes with algebraic structures, but if an S-box does not have an apparent structure, Algorithm\;\ref{alg:synthesis} can be a good candidate for a solution.
We have applied the algorithm to Skipjack\;\cite{skipjack}, KHAZAD\;\cite{khazad}, and DES\;\cite{des} S-boxes, all of which are known not to have efficient classical implementations.
The results are summarized in Table\;\ref{tab:benchmarks}, and the circuits can be found from\;\cite{code}.
For DES (which is an abbreviation for data encryption standard), there are eight different `6-bit input, 4-bit output' S-boxes, and in Table\;\ref{tab:benchmarks} the average values are filled.
Details are given in Appendix\;II.
To our knowledge, none of them have been analyzed before possibly due to the difficulty of handling unstructured permutation maps.
Note that the algorithm is generally applicable to any substitution maps with not too large $n$, giving rise to in-place circuits.

Comparisons with S-boxes that have been previously studied are also given in Appendix\;II.
These S-boxes are either structured or small enough for near optimal synthesis.

\section{Summary and Discussion}\label{sec:6}
In summary, we have developed an algorithm for reversible circuit synthesis based on a transformation-based approach.
The key idea is, in each iteration, the problem size is reduced such that \emph{one bit is completely ruled out} from the remaining process.
In doing so, our primary concern is to minimize the number of Toffoli gates which are treated as the universal gate in quantum computing community in a similar sense to what NAND gate means in classical computing.
Compared with existing algorithms, we have observed that the new algorithm shows good performance in the number of Toffoli gates or QC cost for the functions that are believed to be hard to synthesize, but it is not efficient when the function is structured as expected.

About the time complexity, the algorithm runs in time $O(n 2^{2n})$ which is obviously flawed at least by a factor of $2^n$ even compared with an elementary algorithm such as transforming sequences $(*,*,*,\ldots) \mapsto (1,*,*,\ldots) \mapsto (1,2,*,\ldots) \mapsto \cdots$ by consecutive transpositions.
Although the efficiency in time has not been our concern, it would be a fair question to ask if the same algorithm can run in time $O(n 2^{n})$.
In fact in our analysis, the culprit for the gap is the action of a gate which takes $O(2^n)$ time in our current implementation.
Once a subroutine such as $\mathsf{CONS}$ gives a gate sequence, we do not know the resulting permutation until the \emph{gates are actually applied}.
In comparison with the elementary algorithm, applying transposition immediately gives the resulting permutation, allowing only to count the number of required transpositions to estimate the complexity.
In fact, since the considered elementary gates act only locally, in principle it is possible to bring down its complexity to $\tilde{O}(n2^n)$.
If we are to improve the time complexity, an efficient gate action should be considered such as using tensor network techniques\;\cite{dmrg-tn}.

\bibliographystyle{plain}
\bibliography{reference}

\begin{thebibliography}{10}

\bibitem{code}
{Hochang Lee, Algorithm implementation, Github Repository}.
\newblock 2021.

\bibitem{stabilizer}
Scott Aaronson and Daniel Gottesman.
\newblock Improved simulation of stabilizer circuits.
\newblock {\em Phys. Rev. A}, 70:052328, Nov 2004.

\bibitem{AgrawalJ04}
A.~Agrawal and N.K. Jha.
\newblock Synthesis of reversible logic.
\newblock In {\em Proceedings Design, Automation and Test in Europe Conference
  and Exhibition}, volume~2, pages 1384--1385 Vol.2, 2004.

\bibitem{aes18b}
Mishal Almazrooie, Azman Samsudin, Rosni Abdullah, and Kussay~N. Mutter.
\newblock {Quantum Reversible Circuit of AES-128}.
\newblock {\em Quantum Information Processing}, 17(5):112, Mar 2018.

\bibitem{AMMR13}
Matthew Amy, Dmitri Maslov, Michele Mosca, and Martin Roetteler.
\newblock A meet-in-the-middle algorithm for fast synthesis of depth-optimal
  quantum circuits.
\newblock {\em IEEE Transactions on Computer-Aided Design of Integrated
  Circuits and Systems}, 32(6):818--830, 2013.

\bibitem{RCViewer}
Mona Arabzadeh and Mehdi Saeedi.
\newblock Rcviewer+, version 2.5, 2013.
\newblock 2013.

\bibitem{ACS20}
Pascal Aubry, Sergiu Carpov, and Renaud Sirdey.
\newblock Faster homomorphic encryption is not enough: Improved heuristic for
  multiplicative depth minimization of boolean circuits.
\newblock In Stanislaw Jarecki, editor, {\em Topics in Cryptology -- CT-RSA
  2020}, pages 345--363, Cham, 2020. Springer International Publishing.

\bibitem{barenco95}
Adriano Barenco, Charles~H. Bennett, Richard Cleve, David~P. DiVincenzo, Norman
  Margolus, Peter Shor, Tycho Sleator, John~A. Smolin, and Harald Weinfurter.
\newblock {Elementary gates for quantum computation}.
\newblock {\em Phys. Rev. A}, 52:3457--3467, 1995.

\bibitem{khazad}
PSLM Barreto and Vincent Rijmen.
\newblock {The khazad legacy-level block cipher}.
\newblock {\em Primitive submitted to NESSIE}, 97(106), 2000.

\bibitem{bern09}
Daniel~J Bernstein.
\newblock Cost analysis of hash collisions: Will quantum computers make sharcs
  obsolete.
\newblock {\em Workshop Record of SHARCS'09: Special-purpose Hardware for
  Attacking Cryptographic systems}, 9:105, 2009.

\bibitem{BijChaSan20}
Subodh Bijwe, Amit~Kumar Chauhan, and Somitra~Kumar Sanadhya.
\newblock Quantum search for lightweight block ciphers: {GIFT}, {SKINNY},
  {SATURNIN}.
\newblock Cryptology ePrint Archive, Report 2020/1485, 2020.
\newblock \url{https://eprint.iacr.org/2020/1485}.

\bibitem{aes19}
Xavier Bonnetain, Mar{\'i}a Naya-Plasencia, and Andr{\'e} Schrottenloher.
\newblock {Quantum Security Analysis of AES}.
\newblock {\em {IACR Transactions on Symmetric Cryptology}}, 2019(2):55--93,
  June 2019.

\bibitem{HWB20}
S.~Bravyi, T.~Yoder, and D.~Maslov.
\newblock Efficient ancilla-free reversible and quantum circuits for the hidden
  weighted bit function.
\newblock {\em IEEE Transactions on Computers}, 71(01):1170--1180, apr 2021.

\bibitem{clif-6}
Sergey Bravyi, Joseph~A. Latone, and Dmitri Maslov.
\newblock 6-qubit optimal clifford circuits.
\newblock {\em npj Quantum Information}, 8(1):79, Jul 2022.

\bibitem{Bryant91}
R.E. Bryant.
\newblock {On the complexity of VLSI implementations and graph representations
  of boolean functions with application to integer multiplication}.
\newblock {\em IEEE Transactions on Computers}, 40(2):205--213, 1991.

\bibitem{aria20}
Amit~Kumar Chauhan and Somitra~Kumar Sanadhya.
\newblock {{Quantum resource estimates of Grover's key search on ARIA}}.
\newblock In Lejla Batina, Stjepan Picek, and Mainack Mondal, editors, {\em
  Security, Privacy, and Applied Cryptography Engineering}, pages 238--258,
  Cham, 2020. Springer International Publishing.

\bibitem{ChunBC23}
Matthew Chun, Anubhab Baksi, and Anupam Chattopadhyay.
\newblock {DORCIS:} depth optimized quantum implementation of substitution
  boxes.
\newblock {\em {IACR} Cryptol. ePrint Arch.}, page 286, 2023.

\bibitem{DKRFM21}
Edinel{\c{c}}o Dalcumune, Luis Antonio~Brasil Kowada, Andr{\'{e}}
  da~Cunha~Ribeiro, Celina Miraglia~Herrera de~Figueiredo, and Franklin
  de~Lima~Marquezino.
\newblock A reversible circuit synthesis algorithm with progressive increase of
  controls in generalized toffoli gates.
\newblock {\em J. Univers. Comput. Sci.}, 27(6):544--563, 2021.

\bibitem{routing}
William~James Dally and Brian~Patrick Towles.
\newblock {\em {Principles and practices of interconnection networks}}.
\newblock Morgan Kaufmann Publishers Inc., San Francisco, CA, USA, 2004.

\bibitem{DasuBSC19}
Vishnu~Asutosh Dasu, Anubhab Baksi, Sumanta Sarkar, and Anupam Chattopadhyay.
\newblock {LIGHTER-R:} optimized reversible circuit implementation for sboxes.
\newblock In {\em 32nd {IEEE} International System-on-Chip Conference, {SOCC}
  2019, Singapore, September 3-6, 2019}, pages 260--265. {IEEE}, 2019.

\bibitem{rev-page2}
Maslov Dmitri.
\newblock Reversible logic synthesis benchmarks page.
\newblock 2009.

\bibitem{draper06}
Thomas~G. Draper, Samuel~A. Kutin, Eric~M. Rains, and Krysta~M. Svore.
\newblock {A logarithmic-depth quantum carry-lookahead adder}.
\newblock {\em Quantum Info. Comput.}, 6(4):351–369, July 2006.

\bibitem{fazel07}
K.~Fazel, M.~A. Thornton, and J.~E. Rice.
\newblock Esop-based toffoli gate cascade generation.
\newblock In {\em 2007 IEEE Pacific Rim Conference on Communications, Computers
  and Signal Processing}, pages 206--209, 2007.

\bibitem{perm10}
O.~{Golubitsky}, S.~M. {Falconer}, and D.~{Maslov}.
\newblock {Synthesis of the optimal 4-bit reversible circuits}.
\newblock In {\em Design Automation Conference}, pages 653--656, 2010.

\bibitem{perm12}
Oleg Golubitsky and Dmitri Maslov.
\newblock A study of optimal 4-bit reversible toffoli circuits and their
  synthesis.
\newblock {\em IEEE Transactions on Computers}, 61(9):1341--1353, 2012.

\bibitem{aes16}
Markus Grassl, Brandon Langenberg, Martin Roetteler, and Rainer Steinwandt.
\newblock {Applying \textrm{G}rover's algorithm to \textrm{AES}: quantum
  resource estimates}.
\newblock In {\em PQCrypto 2016}, pages 29--43, 2016.

\bibitem{grover97}
Lov~K. Grover.
\newblock {Quantum mechanics helps in searching for a needle in a haystack}.
\newblock {\em Phys. Rev. Lett.}, 79:325--328, 1997.

\bibitem{gupta06}
P.~Gupta, A.~Agrawal, and N.K. Jha.
\newblock An algorithm for synthesis of reversible logic circuits.
\newblock {\em IEEE Transactions on Computer-Aided Design of Integrated
  Circuits and Systems}, 25(11):2317--2330, 2006.

\bibitem{he17}
Yong He, Ming-Xing Luo, E.~Zhang, Hong-Ke Wang, and Xiao-Feng Wang.
\newblock {Decompositions of n-qubit Toffoli gates with linear circuit
  complexity}.
\newblock {\em International Journal of Theoretical Physics}, 56(7):2350--2361,
  Jul 2017.

\bibitem{JKES20}
Kyoungbae Jang, Hyunjun Kim, Siwoo Eum, and Hwajeong Seo.
\newblock Grover on {GIFT}.
\newblock Cryptology ePrint Archive, Report 2020/1405, 2020.
\newblock \url{https://eprint.iacr.org/2020/1405}.

\bibitem{JBBSC22}
Kyungbae Jang, Anubhab Baksi, Jakub Breier, Hwajeong Seo, and Anupam
  Chattopadhyay.
\newblock Quantum implementation and analysis of {DEFAULT}.
\newblock {\em Cryptography and Communications}, pages 1--17, 2023.

\bibitem{JSK+21}
Kyungbae Jang, Gyeongju Song, Hyeokdong Kwon, Siwoo Uhm, Hyunji Kim, Wai-Kong
  Lee, and Hwajeong Seo.
\newblock Grover on pipo.
\newblock {\em Electronics}, 10(10):1194, 2021.

\bibitem{aes20b}
Samuel Jaques, Michael Naehrig, Martin Roetteler, and Fernando Virdia.
\newblock {{Implementing Grover oracles for quantum key search on AES and
  LowMC}}.
\newblock In Anne Canteaut and Yuval Ishai, editors, {\em Advances in
  Cryptology -- EUROCRYPT 2020}, pages 280--310, Cham, 2020. Springer
  International Publishing.

\bibitem{KaplanLLN16}
Marc Kaplan, Ga{\"{e}}tan Leurent, Anthony Leverrier, and Mar{\'{\i}}a
  Naya{-}Plasencia.
\newblock Breaking symmetric cryptosystems using quantum period finding.
\newblock In Matthew Robshaw and Jonathan Katz, editors, {\em Advances in
  Cryptology - {CRYPTO} 2016 - 36th Annual International Cryptology Conference,
  Santa Barbara, CA, USA, August 14-18, 2016, Proceedings, Part {II}}, volume
  9815 of {\em Lecture Notes in Computer Science}, pages 207--237. Springer,
  2016.

\bibitem{Kerntopf04}
Pawel Kerntopf.
\newblock A new heuristic algorithm for reversible logic synthesis.
\newblock In {\em Proceedings of the 41st Annual Design Automation Conference},
  DAC '04, page 834–837, New York, NY, USA, 2004. Association for Computing
  Machinery.

\bibitem{aes18}
Panjin Kim, Daewan Han, and Kyung~Chul Jeong.
\newblock {Time-space complexity of quantum search algorithms in symmetric
  cryptanalysis: applying to AES and SHA-2}.
\newblock {\em Quantum Information Processing}, 17(12):339, Oct 2018.

\bibitem{optimal-clif-linear}
Vadym Kliuchnikov and Dmitri Maslov.
\newblock Optimization of clifford circuits.
\newblock {\em Phys. Rev. A}, 88:052307, Nov 2013.

\bibitem{aes20}
B.~{Langenberg}, H.~{Pham}, and R.~{Steinwandt}.
\newblock {Reducing the cost of implementing the advanced encryption standard
  as a quantum circuit}.
\newblock {\em IEEE Transactions on Quantum Engineering}, 1:1--12, 2020.

\bibitem{perm08}
Z.~{Li}, H.~{Chen}, B.~{Xu}, X.~{Song}, and X.~{Xue}.
\newblock {An algorithm for synthesis of optimal 3-qubit reversible circuits
  based on bit operation}.
\newblock In {\em 2008 Second International Conference on Genetic and
  Evolutionary Computing}, pages 455--458, 2008.

\bibitem{LXXZZ23}
Da~Lin, Zejun Xiang, Runqing Xu, Xiangyong Zeng, and Shasha Zhang.
\newblock Quantum circuit implementations of {SM4} block cipher based on
  different gate sets.
\newblock {\em Quantum Inf. Process.}, 22(7):282, 2023.

\bibitem{MaslovDM07}
Dmitri Maslov, Gerhard~W. Dueck, and D.~Michael Miller.
\newblock Techniques for the synthesis of reversible toffoli networks.
\newblock {\em {ACM} Trans. Design Autom. Electr. Syst.}, 12(4):42, 2007.

\bibitem{miller04}
D~Michael Miller, Gerhard~W Dueck, and Dmitri Maslov.
\newblock A synthesis method for mvl reversible logic [multiple value logic].
\newblock In {\em Proceedings. 34th international symposium on multiple-valued
  logic}, pages 74--80. IEEE, 2004.

\bibitem{MMD03}
D.~Michael Miller, Dmitri Maslov, and Gerhard~W. Dueck.
\newblock {A transformation based algorithm for reversible logic synthesis}.
\newblock In {\em Proceedings of the 40th Annual Design Automation Conference},
  DAC '03, page 318–323, New York, NY, USA, 2003. Association for Computing
  Machinery.

\bibitem{MillerS12}
D.~Michael Miller and Zahra Sasanian.
\newblock Recent developments on mapping reversible circuits to quantum gate
  libraries.
\newblock In {\em International Symposium on Electronic System Design, ISEDs
  2012, Kolkata, India, December 19-22, 2012}, pages 17--22. {IEEE}, 2012.

\bibitem{QMDD}
D.M. Miller and M.A. Thornton.
\newblock Qmdd: A decision diagram structure for reversible and quantum
  circuits.
\newblock In {\em 36th International Symposium on Multiple-Valued Logic
  (ISMVL'06)}, pages 30--30, 2006.

\bibitem{BVMS04}
Mikko M\"ott\"onen, Juha~J. Vartiainen, Ville Bergholm, and Martti~M. Salomaa.
\newblock {Quantum circuits for general multiqubit gates}.
\newblock {\em Phys. Rev. Lett.}, 93:130502, Sep 2004.

\bibitem{des}
{National Bureau of Standards}.
\newblock {Data Encryption Standard}.
\newblock {\em Fips publication}, (46), 1977.

\bibitem{bookchuang}
Michael~A. Nielsen and Isaac~L. Chuang.
\newblock {\em {Quantum computation and quantum information: 10th anniversary
  edition}}.
\newblock Cambridge University Press, 2010.

\bibitem{skipjack}
{NIST}.
\newblock {Skipjack and KEA algorithm specifications, version 2.0}.
\newblock May 1998.

\bibitem{nist-aes}
NIST.
\newblock {\em {Advanced Encryption Standard}}.
\newblock FIPS PUB 197, 2001.

\bibitem{OJBS23}
Yujin Oh, Kyungbae Jang, Anubhab Baksi, and Hwajeong Seo.
\newblock Depth-optimized implementation of ascon quantum circuit.
\newblock Cryptology ePrint Archive, Paper 2023/1030, 2023.
\newblock \url{https://eprint.iacr.org/2023/1030}.

\bibitem{optimal-linear}
Ketan~N. Patel, Igor~L. Markov, and John~P. Hayes.
\newblock Optimal synthesis of linear reversible circuits.
\newblock {\em Quantum Info. Comput.}, 8(3):282–294, mar 2008.

\bibitem{PhabLS22}
Luca Phab, St{\'{e}}phane Louise, and Renaud Sirdey.
\newblock A first attempt at cryptanalyzing a (toy) block cipher by means of
  {QAOA}.
\newblock In Derek Groen, Cl{\'{e}}lia de~Mulatier, Maciej Paszynski,
  Valeria~V. Krzhizhanovskaya, Jack~J. Dongarra, and Peter M.~A. Sloot,
  editors, {\em Computational Science - {ICCS} 2022 - 22nd International
  Conference, London, UK, June 21-23, 2022, Proceedings, Part {IV}}, volume
  13353 of {\em Lecture Notes in Computer Science}, pages 218--232. Springer,
  2022.

\bibitem{perm06}
Aditya~K. Prasad, Vivek~V. Shende, Igor~L. Markov, John~P. Hayes, and Ketan~N.
  Patel.
\newblock Data structures and algorithms for simplifying reversible circuits.
\newblock {\em J. Emerg. Technol. Comput. Syst.}, 2(4):277–293, oct 2006.

\bibitem{RahPau21}
Mostafizar Rahman and Goutam Paul.
\newblock Grover on present: Quantum resource estimation.
\newblock Cryptology ePrint Archive, Report 2021/1655, 2021.
\newblock \url{https://eprint.iacr.org/2021/1655}.

\bibitem{rev-page}
RevLib.
\newblock An online resources for reversible functions and circuits.
\newblock 2008.

\bibitem{Sae08}
M.~Saeedi.
\newblock {QDA reversible benchmarks}, 2008.
\newblock http://ceit.aut.ac.ir/qda/benchmarks.htm.

\bibitem{SAM11}
Mehdi Saeedi, Mona Arabzadeh, Morteza~Saheb Zamani, and Mehdi Sedighi.
\newblock {Block-based quantum-logic synthesis}.
\newblock {\em Quantum Info. Comput.}, 11(3):262–277, March 2011.

\bibitem{RCT-review}
Mehdi Saeedi and Igor~L. Markov.
\newblock Synthesis and optimization of reversible circuits - a survey.
\newblock volume~45, New York, NY, USA, 2013. Association for Computing
  Machinery.

\bibitem{SZSS10}
Mehdi Saeedi, Morteza~Saheb Zamani, Mehdi Sedighi, and Zahra Sasanian.
\newblock {Reversible circuit synthesis using a cycle-based approach}.
\newblock {\em J. Emerg. Technol. Comput. Syst.}, 6(4), December 2010.

\bibitem{SZM10}
{Saeedi, Mehdi and Zamani, Morteza Saheb and Sedighi, Mehdi and Sasanian,
  Zahra}.
\newblock {Synthesis of reversible circuit using cycle-based approach}.
\newblock {\em J. Emerg. Technol. Comput. Syst}, 6(4):13, 2010.

\bibitem{Sasao93}
Tsutomu Sasao.
\newblock {\em Logic synthesis and optimization}, volume~2.
\newblock Springer, 1993.

\bibitem{dmrg-tn}
Ulrich Schollwöck.
\newblock {{The density-matrix renormalization group in the age of matrix
  product states}}.
\newblock {\em Annals of Physics}, 326(1):96 -- 192, 2011.
\newblock January 2011 Special Issue.

\bibitem{lut13}
A.~{Shafaei}, M.~{Saeedi}, and M.~{Pedram}.
\newblock {Reversible logic synthesis of k-input, m-output lookup tables}.
\newblock In {\em {2013 Design, Automation Test in Europe Conference Exhibition
  (DATE)}}, pages 1235--1240, 2013.

\bibitem{lowerbound}
V.V. Shende, A.K. Prasad, I.L. Markov, and J.P. Hayes.
\newblock Synthesis of reversible logic circuits.
\newblock {\em IEEE Transactions on Computer-Aided Design of Integrated
  Circuits and Systems}, 22(6):710--722, 2003.

\bibitem{SJKESKLS21}
Gyeongju Song, Kyungbae Jang, Hyunjun Kim, Siwoo Eum, Minjoo Sim, Hyunji Kim,
  Wai-Kong Lee, and Hwajeong Seo.
\newblock Grover on {SPEEDY}.
\newblock Cryptology ePrint Archive, Report 2021/1211, 2021.
\newblock \url{https://eprint.iacr.org/2021/1211}.

\bibitem{Tof80}
Tommaso Toffoli.
\newblock {{Reversible computing}}.
\newblock In Jaco de~Bakker and Jan van Leeuwen, editors, {\em Automata,
  Languages and Programming}, pages 632--644, Berlin, Heidelberg, 1980.
  Springer Berlin Heidelberg.

\bibitem{bookJohn}
George~F. Viamontes, Igor~L. Markov, and John~P. Hayes.
\newblock {\em {Quantum Circuit Simulation}}.
\newblock Springer Publishing Company, 2009.

\bibitem{VosR08}
Alexis~De Vos and Yvan~Van Rentergem.
\newblock Young subgroups for reversible computers.
\newblock {\em Adv. Math. Commun.}, 2(2):183--200, 2008.

\bibitem{WilleSPD12}
Robert Wille, Mathias Soeken, Nils Przigoda, and Rolf Drechsler.
\newblock Exact synthesis of toffoli gate circuits with negative control lines.
\newblock In D.~Michael Miller and Vincent~C. Gaudet, editors, {\em 42nd {IEEE}
  International Symposium on Multiple-Valued Logic, {ISMVL} 2012, Victoria, BC,
  Canada, May 14-16, 2012}, pages 69--74. {IEEE} Computer Society, 2012.

\bibitem{perm07}
Guowu Yang, Xiaoyu Song, William N.~N. Hung, and Marek~A. Perkowski.
\newblock {Bi-directional synthesis of 4-bit reversible circuits}.
\newblock {\em The Computer Journal}, 51(2):207--215, 07 2007.

\bibitem{Zak16}
Dmitry~V. Zakablukov.
\newblock Application of permutation group theory in reversible logic
  synthesis.
\newblock In Simon Devitt and Ivan Lanese, editors, {\em Reversible
  Computation}, pages 223--238, Cham, 2016. Springer International Publishing.

\bibitem{zalka00}
Christof Zalka.
\newblock {Using Grover's quantum algorithm for searching actual databases}.
\newblock {\em Phys. Rev. A}, 62:052305, Oct 2000.

\bibitem{whaley04}
Jun Zhang, Jiri Vala, Shankar Sastry, and K.~Birgitta Whaley.
\newblock Optimal quantum circuit synthesis from controlled-unitary gates.
\newblock {\em Phys. Rev. A}, 69:042309, Apr 2004.

\bibitem{aes20c}
Jian Zou, Zihao Wei, Siwei Sun, Ximeng Liu, and Wenling Wu.
\newblock {{Quantum circuit implementations of AES with fewer qubits}}.
\newblock In Shiho Moriai and Huaxiong Wang, editors, {\em Advances in
  Cryptology -- ASIACRYPT 2020}, pages 697--726, Cham, 2020. Springer
  International Publishing.

\end{thebibliography}

\newpage
\appendix
\section{An Example for $\mathsf{PICK}$, $\mathsf{CONS}$ and $\mathsf{ALLOC}$}\label{app:subroutines}
Below we show a 4-bit example of how the subroutines work.
To make it simple, the permutation consists only of row numbers in normal positions.
Consider the following permutation:
\begin{align*}
(\texttt{0},
\texttt{1},
\texttt{2},
\texttt{11},
\texttt{12},
\texttt{3},
\texttt{10},
\texttt{5},
\texttt{4},
\texttt{15},
\texttt{14},
\colorbox[rgb]{0.7,0.7,0.7}{\makebox(0,6){\texttt{7}}},
\colorbox[rgb]{0.7,0.7,0.7}{\makebox(0,6){\texttt{6}}},
\texttt{9},
\texttt{8},
\texttt{13}).
\end{align*}
First, $\mathsf{PICK}$ picks an eligible row pair according to Proposition\;\ref{prop:pick}.
According to the proposition, there exists at least one pair of relevant row numbers located in the right half.
In this example, \texttt{6} and \texttt{7} are the numbers (not the only eligible pair, but the pair with the smallest row numbers is picked for simplicity).
Notice that \texttt{6} and \texttt{7} do not form a block yet, as the two numbers occupy different block-wise positions.
Next, the pair is passed on to $\mathsf{CONS}$ for construction.
The construction rule follows the proof of Lemma\;\ref{lem:construction}.
There already exists a block in the first block-wise position, thus we have $l=1$ (Fig.\;\ref{fig:k-example}).
We want to put a new block at $i=1$ block-wise position, and thus we have $m=2$ (Eq.\;\eqref{eq:m}).
The proof of Lemma\;\ref{lem:construction} describes how to construct a block with only one $C^{m}\!X$ gate which is in this case $C^2 \! X$ gate.
Indeed, $C\!X_{23}$, $X_{4}$, $C\!X_{142}$, $X_{4}$ gates are applied in order leading to,
\begin{align*}
(\texttt{0},
\texttt{1},
\texttt{2},
\texttt{11},
\texttt{10},
\texttt{5},
\texttt{12},
\texttt{3},
\texttt{8},
\texttt{15},
\colorbox[rgb]{0.7,0.7,0.7}{\makebox(9,6){\texttt{6},\texttt{7}}},
\texttt{4},
\texttt{13},
\texttt{14},
\texttt{9}).
\end{align*}
Now, the block is passed on to $\mathsf{ALLOC}$ for the left-allocation.
The rule is as described in the proof of Lemma\;\ref{lem:allocation}.
As mentioned earlier, we want to put a block at $i=1$ and we have $m=2$.
Since $i' = i - h_n(m-1)/2 = 1$ in Lemma\;\ref{lem:allocation}, $i'$ as a bit string is $0001$, and therefore its Hamming weight is 1, telling us that the allocation of the constructed block does not require Toffoli gate (since $C^{HW(\v i')}\!X$ gate is only a CNOT gate now).
Indeed, $C\!X_{13}$ gate is applied leading to,
\begin{align*}
(\colorbox[rgb]{0.7,0.7,0.7}{\makebox(9,6){\texttt{0},\texttt{1}}},
\colorbox[rgb]{0.7,0.7,0.7}{\makebox(9,6){\texttt{6},\texttt{7}}},
\texttt{10},
\texttt{5},
\texttt{14},
\texttt{9},
\texttt{8},
\texttt{15},
\texttt{2},
\texttt{11},
\texttt{4},
\texttt{13},
\texttt{12},
\texttt{3}).
\end{align*}
Repeating the procedure, the others blocks are constructed and allocated as follows:
\begin{align*}
&(\colorbox[rgb]{0.7,0.7,0.7}{\makebox(9,6){\texttt{0},\texttt{1}}},
\colorbox[rgb]{0.7,0.7,0.7}{\makebox(9,6){\texttt{6},\texttt{7}}},
\colorbox[rgb]{0.7,0.7,0.7}{\makebox(19,6){\texttt{10},\texttt{11}}},
\texttt{14},
\texttt{15},
\texttt{12},
\texttt{3},
\texttt{4},
\texttt{13},
\texttt{2},
\texttt{5},
\texttt{8},
\texttt{9})
\\
\mapsto&
(\colorbox[rgb]{0.7,0.7,0.7}{\makebox(9,6){\texttt{0},\texttt{1}}},
\colorbox[rgb]{0.7,0.7,0.7}{\makebox(9,6){\texttt{6},\texttt{7}}},
\colorbox[rgb]{0.7,0.7,0.7}{\makebox(19,6){\texttt{10},\texttt{11}}},
\colorbox[rgb]{0.7,0.7,0.7}{\makebox(19,6){\texttt{14},\texttt{15}}},
\texttt{12},
\texttt{3},
\texttt{4},
\texttt{13},
\texttt{2},
\texttt{5},
\texttt{8},
\texttt{9})
\\
\mapsto&
(\colorbox[rgb]{0.7,0.7,0.7}{\makebox(19,6){\texttt{10},\texttt{11}}},
\colorbox[rgb]{0.7,0.7,0.7}{\makebox(19,6){\texttt{14},\texttt{15}}},
\colorbox[rgb]{0.7,0.7,0.7}{\makebox(9,6){\texttt{0},\texttt{1}}},
\colorbox[rgb]{0.7,0.7,0.7}{\makebox(9,6){\texttt{6},\texttt{7}}},
\colorbox[rgb]{0.7,0.7,0.7}{\makebox(9,6){\texttt{2},\texttt{3}}},
\texttt{8},
\texttt{13},
\texttt{12},
\texttt{5},
\texttt{4},
\texttt{9})
\\
\mapsto&
(\colorbox[rgb]{0.7,0.7,0.7}{\makebox(19,6){\texttt{10},\texttt{11}}},
\colorbox[rgb]{0.7,0.7,0.7}{\makebox(9,6){\texttt{0},\texttt{1}}},
\colorbox[rgb]{0.7,0.7,0.7}{\makebox(9,6){\texttt{6},\texttt{7}}},
\colorbox[rgb]{0.7,0.7,0.7}{\makebox(19,6){\texttt{14},\texttt{15}}},
\colorbox[rgb]{0.7,0.7,0.7}{\makebox(9,6){\texttt{2},\texttt{3}}},
\colorbox[rgb]{0.7,0.7,0.7}{\makebox(9,6){\texttt{4},\texttt{5}}},
\texttt{12},
\texttt{9},
\texttt{8},
\texttt{13})
\\
\mapsto&
(\colorbox[rgb]{0.7,0.7,0.7}{\makebox(9,6){\texttt{0},\texttt{1}}},
\colorbox[rgb]{0.7,0.7,0.7}{\makebox(19,6){\texttt{10},\texttt{11}}},
\colorbox[rgb]{0.7,0.7,0.7}{\makebox(19,6){\texttt{14},\texttt{15}}},
\colorbox[rgb]{0.7,0.7,0.7}{\makebox(9,6){\texttt{6},\texttt{7}}},
\colorbox[rgb]{0.7,0.7,0.7}{\makebox(9,6){\texttt{4},\texttt{5}}},
\colorbox[rgb]{0.7,0.7,0.7}{\makebox(9,6){\texttt{2},\texttt{3}}},
\colorbox[rgb]{0.7,0.7,0.7}{\makebox(9,6){\texttt{8},\texttt{9}}},
\colorbox[rgb]{0.7,0.7,0.7}{\makebox(19,6){\texttt{12},\texttt{13}}})
\end{align*}

The resulting permutation can be understood as a three-bit permutation $(0,\!5,\!7,\!3,\!2,\!1,\!4,\!6)$.
The algorithms $\mathcal{A}_{\rm mix}$ and $\mathcal{A}_{\rm pre}$ transform it into $(3,\!2,\!1,\!7,\!4,\!5,\!6,\!0)$ and repeating the procedures,
\begin{align*}
&(\colorbox[rgb]{0.7,0.7,0.7}{\makebox(9,6){\texttt{4},\texttt{5}}},
\texttt{6},
\texttt{0},
\texttt{3},
\texttt{2},
\texttt{1},
\texttt{7})
\\
~\mapsto~&
(\colorbox[rgb]{0.7,0.7,0.7}{\makebox(9,6){\texttt{4},\texttt{5}}},
\colorbox[rgb]{0.7,0.7,0.7}{\makebox(9,6){\texttt{6},\texttt{7}}},
\texttt{3},
\texttt{2},
\texttt{1},
\texttt{0})
\\
~\mapsto~&
(	\colorbox[rgb]{0.7,0.7,0.7}{\makebox(9,6){\texttt{6},\texttt{7}}},
\colorbox[rgb]{0.7,0.7,0.7}{\makebox(9,6){\texttt{4},\texttt{5}}},	
\colorbox[rgb]{0.7,0.7,0.7}{\makebox(9,6){\texttt{1},\texttt{0}}},
\colorbox[rgb]{0.7,0.7,0.7}{\makebox(9,6){\texttt{3},\texttt{2}}})
\\
\overset{C\!X_{12}}{~\mapsto~}&
(	\colorbox[rgb]{0.7,0.7,0.7}{\makebox(9,6){\texttt{6},\texttt{7}}},
\colorbox[rgb]{0.7,0.7,0.7}{\makebox(9,6){\texttt{4},\texttt{5}}},	
\colorbox[rgb]{0.7,0.7,0.7}{\makebox(9,6){\texttt{0},\texttt{1}}},
\colorbox[rgb]{0.7,0.7,0.7}{\makebox(9,6){\texttt{2},\texttt{3}}}).
\end{align*}

Again it can be read as a smaller permutation $(3,2,0,1)$.
Applying $X_1$ followed by $C\!X_{12}$ completes the synthesis.

\newpage
\section{Application to cryptography}\label{app:crpytanalysis}
Various aspects of the proposed algorithm applied to quantum cryptography is discussed in this section.

\subsection{Discussion}
\label{sec:cryptanalysis-a}
In quantum cryptography, especially in cryptanalysis, researchers often produce their result in terms of the exact circuit, not in the asymptotic form.
In many cases this entails generating a quantum circuit for the target cipher itself, for example a number of papers that optimize the quantum implementation of AES algorithm can be found in the literature~\cite{aes16,aes18b,aes19,aes20,aes20b,aes20c}.
The proposed algorithm can be an option for such tasks, especially when the nonlinear part of the cipher is to be synthesized since the linear part is known to be buildable without needs for non-Clifford gates nor extra qubits.

There could be various ways to make use of the proposed algorithm, but for now we can think of two main use cases and one promising direction.
The first one is that the attack cost should be minimized for the space.
One caveat is that this case may not be the usual scenario in Grover where reducing the space requirement does not ‘proportionally’ increase the time.
See,~\cite{bern09} for the details.
Therefore this usage is better be avoided in Grover attacks unless other options are limited.
Instead, one may look for the cryptanalysis by using Simon or quantum approximate optimization algorithms~\cite{KaplanLLN16,PhabLS22}.
As such algorithms have studied relatively less than Grover attacks, a study of cryptanalysis by using the proposed algorithm may require an independent project.

The other case the algorithm can be useful is where the nonlinear part of the cipher is hard to quantumly implement.
In less scientific terms, we believe if a nonlinear part is over 5-bit size ‘and’ if the part does not have efficient algebraic structures (one does not know how to simply simulate the function by additions or multiplications), the proposed algorithm can be a good option.
We have indirect clues.
At the time of writing, there exist a number of studies on the quantum implementations of 5-bit or smaller S-boxes~\cite{DasuBSC19,BijChaSan20,JKES20,JSK+21,RahPau21,JBBSC22,ChunBC23,OJBS23}, but not many results are out there for larger than 5-bit ones when the algebraic construction is hindered but not entirely impossible~\cite{SJKESKLS21}.
Related with the issue, it is worth noting that for 5-bit or smaller functions, near optimal algorithms exist which use exhaustive or meet-in-the-middle techniques.
DORCIS algorithm introduced in Section~\ref{appSubSec:Various sboxes} is a good example.

Lastly, it has been pointed out by an anonymous referee that synthesis algorithms can be applied to improve the performance of homomorphic encryptions by optimizing the multiplicative depth of Boolean circuits\cite{ACS20}.
This also deserves an independent study which may potentially find a real-world application.

Right directions for applying the proposed algorithm to cryptography are discussed above, but in this section we only examine the limited applicability and provide the data to the readers mainly for comparisons.
Below we make comparisons with various S-boxes that have been and have not been studied before.



\subsection{Previously studied S-boxes}
\label{appSubSec:Various sboxes}
\begin{table}[H]
	\centering
	\caption{Comparisons of previously studied S-boxes in quantum cryptanalysis. For ARIA, the result is given by Toffoli-depth. For ASCON, the number in the square bracket means the number of garbage qubits generated.}
	\setlength\tabcolsep{2pt}
	\begin{tabular}{c|cccc}
		\hline
		\multirow{2}{*}{Target} & \multicolumn{2}{c}{Previous work} & \multicolumn{2}{c}{This work}\\
		& \#qubits & depth & \#qubits & depth\\\hline
		AES\cite{aes20b} & 137 & 6\;(T) & 15 & 579\;(T)\\
		ARIA\cite{aria20} & 40 & 196\;(Tof) & 11 & 749\;(Tof)\\
		SM4\cite{LXXZZ23} & 13 & 72\;(T) & 15 & 594\;(T)\\
		ASCON\cite{OJBS23} & 15\;[5] & 0\;(T) & 10 & 24\;(T)\\
		DEFAULT$_{\text{Core}}$\cite{ChunBC23} & 8 & 4\;(T) & 8 & 6\;(T)\\
		DEFAULT$_{\text{Layer}}$\cite{ChunBC23} & 8 & 1\;(T) & 8 & 5\;(T)\\\hline
	\end{tabular}
	\label{tab:comparing with DORCIS}
\end{table}
Three 8-bit, one 5-bit, and two 4-bit S-boxes are examined and summarized in Table~\ref{tab:comparing with DORCIS}.
All 8-bit S-boxes (AES, ARIA, SM4) utilize the inverse of a field element in $\mathbb{F}_{2^8}$, enabling an efficient construction.
Also, those three ciphers happen to be the only ones with relatively large (over 5-bit) S-boxes that we can find in the literature so far.\footnote{There exists a 6-bit S-box studied in\cite{SJKESKLS21}, but it does not provide a suitable data format.}

About 4-bit S-boxes, there exist a handful of ciphers previously studied, but we only examine the two in the table.
Note that for these S-boxes the algorithm designed by\cite{ChunBC23} should work better than most heuristic algorithms as it should be near optimal for small S-boxes.

\subsection{Unstructured S-boxes}
\label{app:unstructured-sboxes}
Relatively larger S-boxes ($>5$) without algebraic structures have attracted less attention from the community, possibly due to the factor discussed in Section\;\ref{sec:cryptanalysis-a}.
We choose DES, Skipjack, and KHAZAD for the application.
There are eight different 6-bit input 4-bit output DES S-boxes.
In Table\;\ref{tab:sboxes}, we summarize the results of all DES S-boxes together with Skipjack and KHAZAD.
(In Table\;\ref{tab:benchmarks}, only the averaged values of QC and the number of Toffoli gates are presented.)

\begin{table*}[ht]
	\centering
	\caption{Costs of reversible circuits for various unstructured S-boxes, obtained by the algorithm with depth-3 partial search. DES has eight different S-boxes, where each takes as input a 6-bit string and outputs a 4-bit string. Two bits of garbage are unavoidable in the reversible implementation of the DES S-box.}
	\linespread{1}
	\setlength\tabcolsep{2pt}
	\begin{tabular}{>{\centering}p{1.9cm}| >{\centering}p{0.83cm}| >{\centering}p{0.9cm}| >{\centering}p{0.83cm}| >{\raggedleft}p{1.1cm}| >{\raggedleft}p{1.1cm} | >{\raggedleft}p{1.1cm}| >{\raggedleft}p{1.1cm} | >{\raggedleft}p{1.1cm}| >{\raggedleft}p{1.1cm} | >{\raggedleft}p{1.1cm}| >{\raggedleft}p{1.1cm}}
		\multicolumn{4}{c|}{}&\multicolumn{2}{c|}{$d=0$}&\multicolumn{2}{c|}{$d=1$}&\multicolumn{2}{c|}{$d=2$}&\multicolumn{2}{c}{$d=3$}\tabularnewline\cline{5-12}
		S-box    & {\#in} & {\#out} & {\#grb} & \centering{QC}& \centering{\#TOF} & \centering{QC}& \centering{\#TOF} & \centering{QC}& \centering{\#TOF} & \centering{QC}& \centering{\#TOF}	\tabularnewline \hline
		DES-1    & 6 & 4 & 2 & 946  & 143	& 679	& 93	& 714	& 97	& 721	& 95	\tabularnewline
		DES-2    & 6 & 4 & 2 & 874  & 121	& 763	& 103	& 760	& 101	& 672	& 92	\tabularnewline
		DES-3    & 6 & 4 & 2 & 845  & 123	& 723	& 104	& 723	& 104	& 731	& 104	\tabularnewline
		DES-4    & 6 & 4 & 2 & 874  & 129	& 698	& 97	& 684	& 94	& 684	& 94	\tabularnewline
		DES-5    & 6 & 4 & 2 & 904  & 128	& 721	& 101	& 756	& 102	& 739	& 101	\tabularnewline
		DES-6    & 6 & 4 & 2 & 880  & 128	& 714	& 99	& 718	& 102	& 794	& 112	\tabularnewline
		DES-7    & 6 & 4 & 2 & 879  & 125	& 791	& 109	& 791	& 109	& 718	& 101	\tabularnewline
		DES-8    & 6 & 4 & 2 & 946  & 135	& 776	& 104	& 805	& 112	& 744	& 100	\tabularnewline
		Skipjack & 8 & 8 & 0 & 7553 & 1100	& 5575	& 803	& 5562	& 791	& 5440	& 771	\tabularnewline
		KHAZAD   & 8 & 8 & 0 & 7578 & 1075	& 5568	& 819	& 5411	& 794	& 5126	& 742	\tabularnewline \hline
	\end{tabular}
	\label{tab:sboxes}
\end{table*}

\end{document}